\documentclass[useAMS,usenatbib]{mn2e}
\usepackage{times}
\usepackage{amsmath}
\usepackage{graphicx}
\newcommand{\mpc}{\mbox{$ h^{-1} \rmn{Mpc}$}}
\newcommand{\xigal}{\mbox{$\xi_{\rmn{gal}}$}}
\newcommand{\xidm}{\mbox{$\xi_{\rmn{dm}}$}}
\newcommand{\lya}{\mbox{$\rmn{Ly}\alpha$}}
\newcommand{\galform}{\texttt{GALFORM}}
\newcommand{\llya}{\mbox{$L_{\rmn{Ly}\alpha}$}}						
\newcommand{\flya}{\mbox{$F_{\rmn{Ly}\alpha}$}}						
\newcommand{\lunits}{\mbox{$[\rmn{erg} \ \rmn{s}^{-1} \ \rmn{h}^{-2}]$}}		
\newcommand{\funits}{\mbox{$[\rmn{erg} \ \rmn{s}^{-1} \ \rmn{cm}^{-2}]$}}		
\newcommand{\ewobs}{\mbox{$EW_{\rmn{obs}}$}}						
\newcommand{\arcm}{\mbox{$\rmn{[arcmin]}$}}						
\newcommand{\wtheta}{\mbox{$w(\theta)$}}						
\newcommand{\zreion}{\mbox{$z_{\rm{reion}}$}}						
\title[The clustering of \lya\ emitters]
{The clustering of \lya\ emitters in a $\Lambda$CDM Universe}
\author [A. Orsi et al.]
{Alvaro Orsi,$^{1,2}$\thanks{Email: alvaro.orsi@durham.ac.uk} Cedric G. Lacey,$^1$ Carlton M. Baugh$^1$ and Leopoldo Infante$^2$ \\
$^1$ Institute for Computational Cosmology, Department of Physics, University of Durham, South Road, Durham DH1 3LE \\
$^2$ Departamento de Astronom\'ia y Astrof\'isica, Facultad de F\'isica, Pontificia Universidad Cat\'olica de Chile, Casilla 306, Santiago 22, Chile}
\begin{document}
\maketitle
\begin{abstract}
We combine a semi-analytical model of galaxy formation with a very
large simulation which follows the growth of large scale structure in
a $\Lambda$CDM universe to predict the clustering of \lya\
emitters. We find that the clustering strength of \lya\ emitters has
only a weak dependence on \lya\ luminosity but a strong dependence on
redshift. With increasing redshift, \lya\ emitters trace progressively
rarer, higher density regions of the universe.  Due to the large
volume of the simulation, over 100 times bigger than any previously
used for this application, we can construct mock catalogues of \lya\
emitters and study the sample variance of current and forthcoming
surveys. We find that the number and clustering of \lya\ emitters in
our mock catalogues are in agreement with measurements from current
surveys, but that there is a considerable scatter in these
quantities. We argue that a proposed survey of emitters at $z=8.8$
should be extended significantly in solid angle to allow a robust
measurement of \lya\ emitter clustering.
\end{abstract}


\section{Introduction}
The study of galaxies at high redshifts opens an important window on
the process of galaxy formation and conditions in the early universe.
The detection of populations of galaxies at high redshifts is one of
the great challenges in observational cosmology. Currently three main
observational techniques are used to discover high redshift,
star-forming galaxies: (i) The Lyman-break drop-out technique, in
which a galaxy is imaged in a combination of three or more optical or
near-IR bands. The longer wavelength filters detect emission in the
rest-frame ultraviolet from ongoing star formation, whereas the
shorter wavelength filters sample the Lyman-break feature. Hence, a
Lyman-break galaxy appears blue in one colour and red in the other
\citep{steidel96, steidel99}. By shifting the whole filter set to
longer wavelengths, the Lyman-break feature can be isolated at higher
redshifts; (ii) Sub-millimetre emission, due to dust being heated when
it absorbs starlight \citep{smail97, hughes98}. The bulk of the energy
absorbed by the dust comes from the rest-frame ultra-violet and so the
dust emission is sensitive to the instantaneous star formation rate;
(iii) Ly-$\alpha$ line emission from star forming galaxies, typically
identified using either narrowband imaging \citep{hu98, kudritzki00,
gawiser07, ouchi07} or long-slit spectroscopy of gravitationally
lensed regions \citep{ellis01, santos04, stark07}. The Ly-$\alpha$
emission is driven by the production of Lyman-continuum photons and so
is dependent on the current star formation rate.

The Lyman-break drop-out and sub-millimetre detection methods are more
established than Ly-$\alpha$ emission as a means of identifying
substantial populations of high redshift galaxies. Nevertheless, in
the last few years there have been a number of \lya\ surveys which
have successfully found high redshift galaxies e.g. \citep{hu98,
kudritzki00}. The observational samples have grown in size such that
statistical studies of the properties of \lya\ emitters have now become
possible: for example, the SXDS Survey \citep{ouchi05,ouchi07} has
allowed estimates of the luminosity function (LFs) and clustering of
\lya\ emitters in the redshift range $3<z<6$, and the MUSYC survey
\citep{gronwall07,gawiser07} has also produced clustering measurements
at $z \sim 3$. Furthermore, the highest redshift galaxy ($z=6.96$)
robustly detected to date was found using the \lya\ technique
\citep{iye}.  Taking advantage of the magnification of faint sources
by gravitational lensing, \citet{stark07} reported 6 candidates for
\lya\ emitters in the redshift range $8.7<z<10.2$, but these have yet
to be confirmed. The DAzLE Project \citep{horton} is designed to find
\lya\ emitters at $z=7.73$ and $z=8.78$.  However, the small field of
view of the instrument ($6.83 \arcmin \times 6.83 \arcmin$) makes it
difficult to use to study large scale structure (LSS) at such
redshifts.  On the other hand, the ELVIS Survey \citep{kim07a, kim07b}
would appear to offer a more promising route to study the LSS of very
high redshift galaxies ($z=8.8$).

Despite these observational breakthroughs, predictions of the
properties of star-forming \lya\ emitting galaxies are still in the
relatively early stages of development. Often these calculations
employ crude assumptions about the galaxy formation process to derive
a star formation rate and hence a \lya\ luminosity, or use
hydrodynamical simulations, which, due to the high computational
overhead, study relatively small cosmological volumes. \citet{hai}
made predictions for the escape fraction of \lya\ emission and the
abundance of \lya\ emitters using the Press-Schechter formalism and a
prescription for the dust distribution in galaxies. Radiative transfer
calculations of the escape fraction have been made by \citet{zheng02},
\citet{ahn04} and \citet{verhamme06} for idealized geometries, while
\citet{tasitsiomi06} and \citet{laursen07} applied these calculations
to galaxies taken from cosmological hydrodynamical
simulations. \citet{barton04} and \citet{furlanetto05} calculated the
number density of Ly-$\alpha$ emitters using hydrodynamical
simulations of galaxy formation. Nagamine et~al. (2006, 2008) used
hydrodynamical simulations to predict the abundance and clustering of
\lya\ emitters. The typical computational boxes used in these
calculations are very small ($\sim 10-30 h^{-1}$Mpc), which makes it
impossible to evolve the simulation accurately to $z=0$. Hence, it is
difficult to test if the galaxy formation model adopted produces a
reasonable description of present day galaxies. Furthermore, the small
box size means that reliable clustering predictions can only be
obtained on scales smaller than the typical correlation length of the
galaxy sample. As we will show in this paper, small volumes are
subject to significant fluctuations in clustering amplitude.

The semi-analytical approach to modelling galaxy formation allows us
to make substantial improvements over previous calculations of the
properties of \lya\ emitters. The speed of this technique means that
large populations of galaxies can be followed. The range of
predictions which can be made using semi-analytical models is, in
general, broader than that produced from most hydrodynamical
simulations, so that the model predictions can be compared more
directly with observational results. A key advantage is that the
models can be readily evolved to the present-day, giving us more faith
in the ingredients used; i.e. we can be reassured that the physics
underpinning the predictions presented for a high-redshift population
of galaxies would not result in too many bright/massive galaxies at
the present day.

The first semi-analytical calculation of the properties of \lya\
emitters based on a hierarchical model of galaxy formation was carried
out by \citet{dell1}. This is the model used throughout this work,
which has been shown to be successful in predicting the properties of
\lya\ emitters over a wide range of redshifts. The semi-analytical
model allows us to connect \lya\ emission to other galaxy
properties. \citet{dell2} showed that this model succesfully predicts
the observed \lya\ LFs and equivalent widths (EWs), along with some
fundamental physical properties, such as star formation rates (SFRs),
gas metallicities, and stellar and halo masses. In \citet{kim07a}, we
used the model to make further predictions for the LF of very high
redshift \lya\ emitters and to study the feasibility of current and
forthcoming surveys which aim to detect such high redshift
galaxies. \citet{kobayashi07} developed an independent semi-analytical
model to derive the luminosity functions of \lya\ emitters.

The focus of this paper is to use the model introduced by
\citet{dell1} to study the clustering of high-redshift \lya\ emitting
galaxies and to extend the comparison of model predictions with
current observational data. \citet{dell2} already gave an indirect
prediction of the clustering of \lya\ emitters by studying galaxy bias
as a function of \lya\ luminosity. However, these results depend on an
analytical model for the halo bias \citep{sheth01}, and furthermore
the linear bias assumption breaks down on small scales. Here we will
present an explicit calculation of the clustering of galaxies by
implementing the semi-analytical model on top of a large N-body
simulation of the hierarchical clustering of the dark matter
distribution. This allows us to predict the spatial distribution of
\lya\ emitting galaxies, and to create realistic maps of \lya\
emitters at different redshifts. These maps can be analysed with
simple statistical tools to quantify the spatial distribution and
clustering of galaxies at high redshifts. The N-body simulation used
in this work is the \textit{Millennium Simulation}, carried out by the
Virgo Consortium \citep{mill}.  The simulation of the spatial
distribution of \lya\ emitters is tested by creating mock catalogues
for different surveys of \lya\ emitting galaxies in the range $3<z<9$.
The clustering of \lya\ emitters in our model is analysed with
correlation functions and halo occupation distributions. Taking
advantage of the large volume of the Millennium simulation, we also
compute the errors expected on correlation function measurements from
various surveys due to cosmic variance.

The outline of this paper is as follows: Section 2 gives a brief
description of the semi-analytical galaxy formation model and
describes how it is combined with the N-body simulation. In Section 3
we establish the range of validity of our simulated galaxy samples by
studying the completeness fractions in the model \lya\ luminosity
functions.  Section 4 gives our predictions for the clustering of
\lya\ emitters in the range $0<z<9$.  In Section 5 we compare our
simulation with recent observational data and we also make predictions
for future measurements (clustering and number counts) expected from
the ELVIS Survey. Finally, Section 6 gives our conclusions.

\section{The Model}

We use the semi-analytical model of galaxy formation, \galform, to
predict the properties of the \lya\ emission of galaxies and their
abundance as a function of redshift. The \galform\ model is fully
described in \citet{cole00} \citep[see also the review by][]{baugh06}
and the variant used here was introduced by \citet{baugh05} \citep[see
also][for a more detailed description]{lacey08}.  The model computes
star formation histories for the whole galaxy population, following
the hierarchical evolution of the host dark matter haloes.

The merger histories of dark matter haloes are calculated using a
Monte Carlo method, following the formalism of the extended Press \&
Schechter theory \citep{press74, lacey93}.  When using Monte Carlo
merger trees, the mass resolution of dark matter haloes can be
arbitrarily high, since the whole of the computer memory can be
devoted to one tree rather than a population of trees. In contrast,
N-body merger trees are constrained by a finite mass resolution due to
the particle mass, which is usually poorer than that typically adopted
for Monte Carlo merger trees. Discrepancies between the model
predictions obtained with Monte Carlo trees and  those extracted
from a simulation only become evident fainter than some luminosity
which is set by the mass resolution of the N-body trees, as we will
see in the next section \citep[see also ][]{helly03}.

\begin{figure*}
\centering
\includegraphics[width=8.5cm]{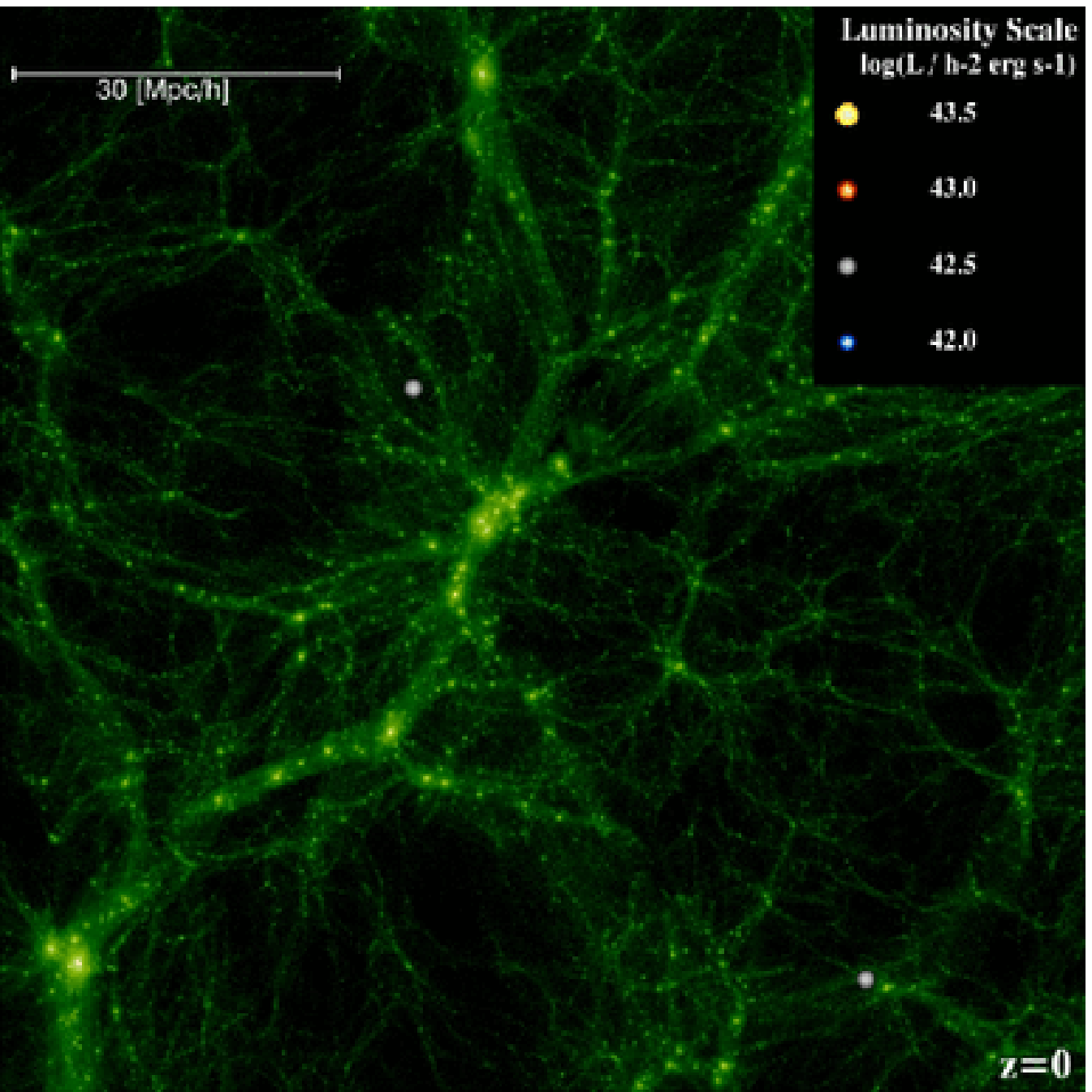}
\includegraphics[width=8.5cm]{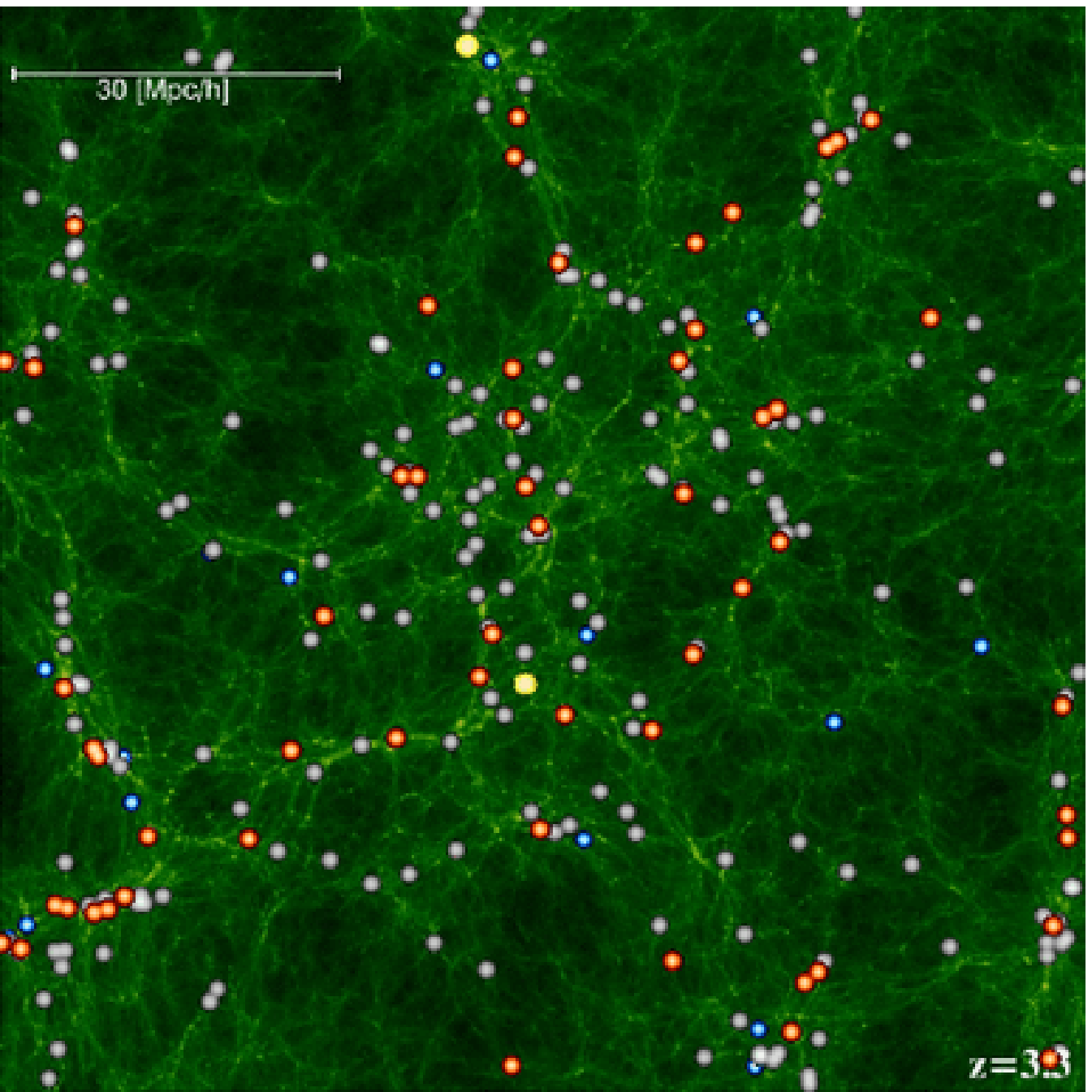}
\includegraphics[width=8.5cm]{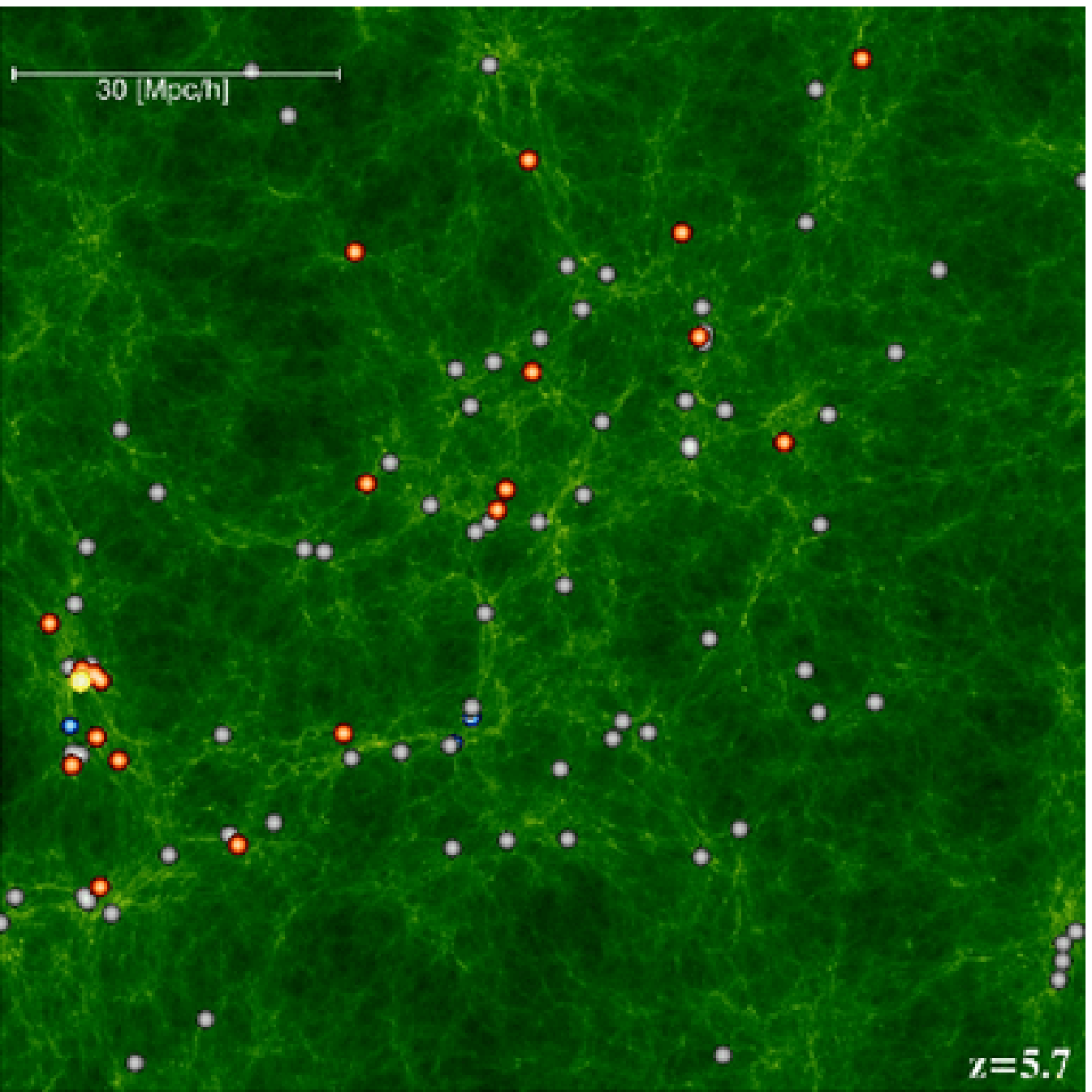}
\includegraphics[width=8.5cm]{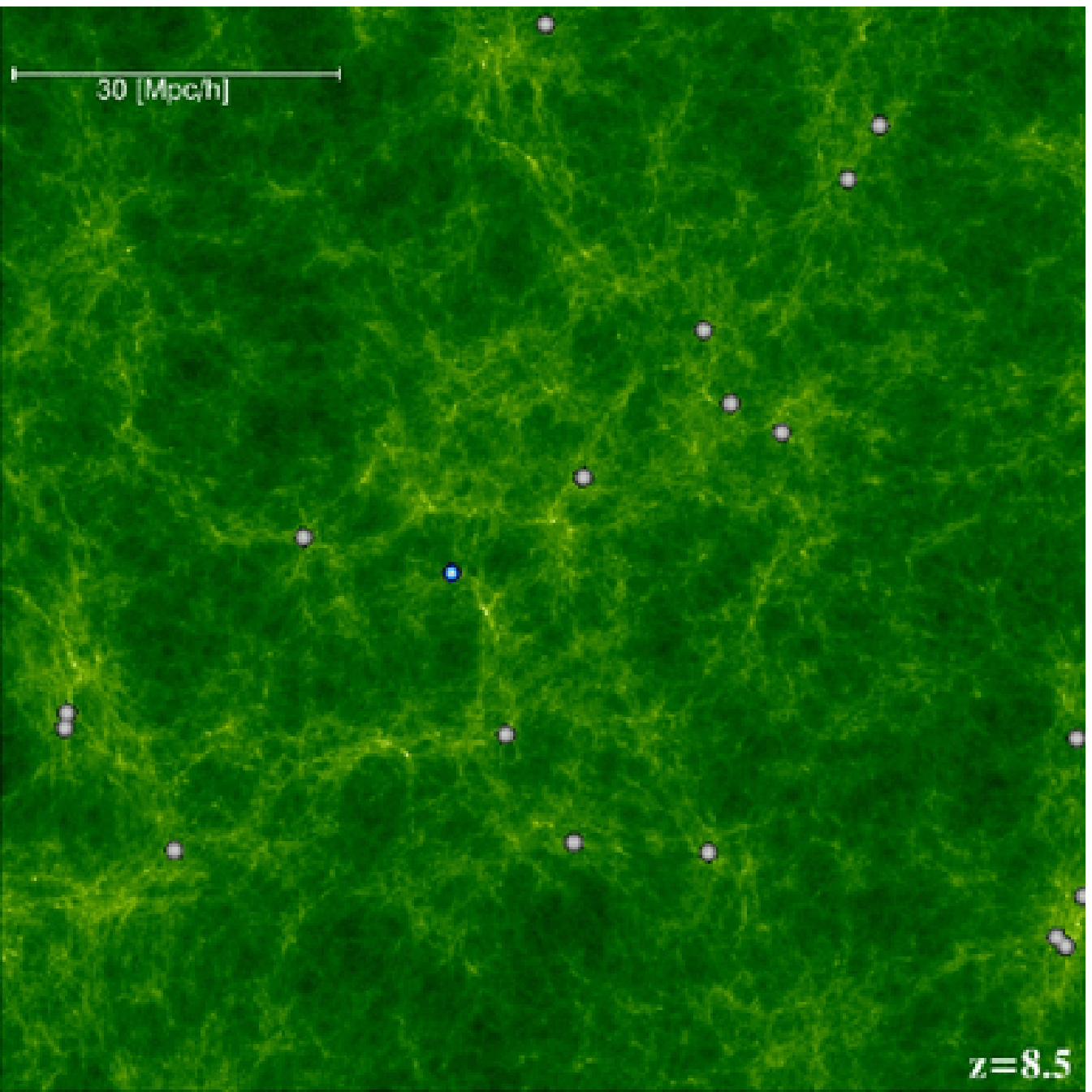}
\caption{ The spatial distribution of \lya\ emitting galaxies (coloured
circles) in a slice from the Millennium simulation, with the dark
matter distribution in green. The four panels are for redshifts in the
range $0<z<8.5$, as indicated in each panel. The colour of the 
circles changes with the \lya\ luminosity of the galaxies, as shown
in the key in the upper-right corner of the first panel. Only galaxies
brighter than $\log(\llya\lunits) = 42.2$ are plotted. Each image
covers a square region $100\times 100 \mpc^2$ across and having a
depth of 10\mpc, which is less than one thousandth the volume of the
full simulation box.}
\label{fig.sample}
\end{figure*}

A critical assumption of the Baugh et al. model is that stars formed
in starbursts have a top-heavy initial mass function (IMF), where the
IMF is given by $\rm{d}N/\rm{d} \ln (m) \propto m^{-x}$ and
$x=0$. Stars formed quiescently in discs have a solar neighbourhood
IMF, with the form proposed by \citet{ken83}: $x=0.4$ for $m < 1
M_\odot$ and $x = 1.5$ for $m> 1 M_\odot$. Both IMFs cover the mass
range $0.15 M_\odot < m < 125 M_\odot$. Within the framework of
$\Lambda$CDM, Baugh et~al.  argued that the top-heavy IMF is essential
to match the counts and redshift distribution of galaxies detected
through their sub-millimetre emission, whilst retaining the match to
galaxy properties in the local Universe, such as the optical and
far-IR luminosity functions and galaxy gas fractions and
metallicities. \citet{nagashima05a,nagashima05b} showed that such a
top-heavy IMF also results in predictions for the metal abundances in
the intra-cluster medium and in elliptical galaxies in much better
agreement with observations. \citet{lacey08} showed that the same
model predicts galaxy evolution in the IR in good agreement with
observations from {\em Spitzer}, and also discussed independent
observational evidence for a top-heavy IMF.

Reionization is assumed in our model to occur at $\zreion=10$
\citep{kogut03,dunkley08}. The photoionization of the intergalactic
medium (IGM) is assumed to suppress the collapse and cooling of gas in
haloes with circular velocities $V_c < 60 \rm{km/s}$ at redshifts
$z<\zreion$ \citep{benson02}.  Recent calculations
\citep{hoeft06,okamoto08} imply that the above parameter values
overstate the impact of photoionization on gas cooling, and suggest
that photoionization only affects smaller haloes with $V_c \la 30 {\rm
km/s}$. Our model predictions, for the range of \lya\ luminosities we
consider in this paper, are not significantly affected by adopting the
lower $V_c$ cut.  We neglect any attenuation of the \lya\ flux by
propagation through the IGM.  This effect would suppress the observed
\lya\ flux mainly for $z \ga \zreion$, so this should only affect our
results for very high redshifts ($z \ga 10$).

The model used to predict the luminosities and equivalent widths of
the \lya\ galaxies is identical to that described in \citet{dell1,
dell2}. The \lya\ emission is computed by the following procedure: (i)
The integrated stellar spectrum of the galaxy is calculated, based on
its star formation history, including the effects of the distribution
of stellar metallicities, and taking into account the IMFs adopted for
different modes of star formation. (ii) The rate of production of
Lyman continuum (Lyc) photons is computed by integrating over the
stellar spectrum, and assuming that all of these ionizing photons are
absorbed by neutral hydrogen within the galaxy. 
We calculate the fraction of \lya\ photons produced by these 
Lyc photons, assuming Case B recombination \citep{osterbrock89}.

(iii) The observed \lya\ flux depends on the fraction of \lya\ photons which escape from
the galaxy ($f_{\rmn{esc}}$), which is assumed to be constant and
independent of galaxy properties.

Calculating the \lya\ escape fraction from first principles by
following the radiative transfer of the \lya\ photons is very
demanding computationally. A more complete calculation of the escape 
fraction would have into account the structure and kinematic properties 
of the intestellar medium (ISM) \citep{zheng02, ahn04, verhamme06}.
In our model, we adopt the simplest possible approach, which is to
fix the escape fraction, $f_{\rmn{esc}}$, to be the same for each
galaxy, without taking into account its dust properties. This results
in a surprisingly good agreement between the predicted number counts
and luminosity functions of emitters and the available observations at
$3 \la z \la 7$ \citep{dell1,dell2}. \citet{dell1} chose
$f_{\rmn{esc}} = 0.02$ to match the number counts at $z\approx 3$ at a
flux $f \approx 2 \times 10^{-17} {\rm erg cm^{-2} s^{-1}}$.  The same
value is used in this work. This value for the \lya\ escape fraction
seems very small, but is consistent with direct observational
estimates for low redshift galaxies: \citet{Atek08} derive escape
fractions for a sample of nearby star-forming galaxies by combining
measurements of Ly$\alpha$, H$\alpha$ and H$\beta$, and find that most
have escape fractions of 3\% or less.  \citet{dell2} also showed that
if a standard solar neighbourhhod IMF is adopted for all modes of star
formation, then a substantially larger escape fraction would be
required to match the observed counts of \lya\ emitters, and even then
the overall match would not be as quite good as it is when the
top-heavy IMF is used in bursts.

Once we obtain the galaxy properties from the semi-analytical model,
we plant these galaxies into a N-body simulation, in order to add
information about their positions and velocities. The simulation used
here is the \textit{Millennium Simulation} \citep{mill}.  This
simulation adopts concordance values for the parameters of a flat
$\Lambda$CDM model, $\Omega_{\rm{m}} = 0.25$ and $\Omega_{\rm{b}} =
0.045$ for the densities of matter and baryons at $z=0$, $\rmn{h} =
0.73$ for the present-day value of the dimensionless Hubble constant,
$\sigma_8 = 0.9$ for the \textit{rms} linear mass fluctuations in a
sphere of radius $8 \rmn{h}^{-1}$Mpc at $z=0$ and $n=1$ for the slope
of the primordial fluctuation spectrum. The simulation follows
$2160^3$ dark matter particles from $z=127$ to $z=0$ within a cubic
region of comoving length $500\rmn{h}^{-1}\rmn{Mpc}$. The individual
particle mass is $8.6 \times 10^8\rmn{h}^{-1}\rmn{M_{\odot}}$, so the
smallest dark halo which can be resolved has a mass of $2 \times
10^{10}\rmn{h}^{-1}\rmn{M_{\odot}}$. 

Dark matter haloes are identified using a \textit{Friends-Of-Friends}
(FOF) algorithm.  To populate the simulation with galaxies from the
semi-analytical model, we use the same approach as in
\citet{benson00}. First, the position and velocity of the centre of
mass of each halo is recorded, along with the positions and velocities
of a set of randomly selected dark matter particles from each
halo. Second, the list of halo masses is fed into the semi-analytical
model in order to produce a population of galaxies associated with
each halo. Each galaxy is assigned a position and velocity within the
halo. Since the semi-analytical model distinguishes between central
and satellite galaxies, the central galaxy is placed at the centre of
mass of the halo, and any satellite galaxy is placed on one of the
randomly selected halo particles. Once galaxies have been generated,
and positions and velocities have been assigned, it is a simple
process to produce catalogues of galaxies with spatial information and
any desired selection criteria.

The combination of the semi-analytical model with the N-body
simulation is essential to study the detailed clustering of a desired
galaxy population, although the clustering amplitude on large scales
can also be estimated analytically \citep{dell2}. An example of the
output of the simulation is shown in the four images of
Fig.~\ref{fig.sample} which show redshifts $z=0$, $z=3.3$, $z=5.7$ and
$z=8.5$. The dark matter distribution (shown in green) becomes
smoother as we go to higher redshifts, due to the gravitational growth
of structures. As shown in Fig.~\ref{fig.sample}, for this particular
luminosity cut, the number density of \lya\ emitters varies at
different redshifts. As we will show in the next section, these
catalogues at high redshift are not complete at faint luminosities, so
we have to restrict our predictions to brighter luminosities as we go
to higher redshifts.

\section{Luminosity Functions}

The model presented by \citet{dell1, dell2} differs in two main ways
from the one presented in this paper: (i) there is a slight difference
in the values of the cosmological parameters used, and (ii) the
earlier work used a grid of halo masses together with an analytical
halo mass function, rather than the set of haloes from an N-body
simulation. In \S3.1, we investigate the impact of the different
choice of cosmological parameters on the luminosity function of \lya\
emitters, to see if the very good agreement with observational data
obtained by \cite{dell2} is retained on adopting the Millennium
cosmology.  In \S3.2, we assess the completeness of our samples of
\lya\ emitters due to the finite mass resolution of the Millennium
simulation.

\subsection{Comparison of model predictions with observed luminosity functions}
\begin{figure}
 \centering
\includegraphics[width=7cm,angle=90]{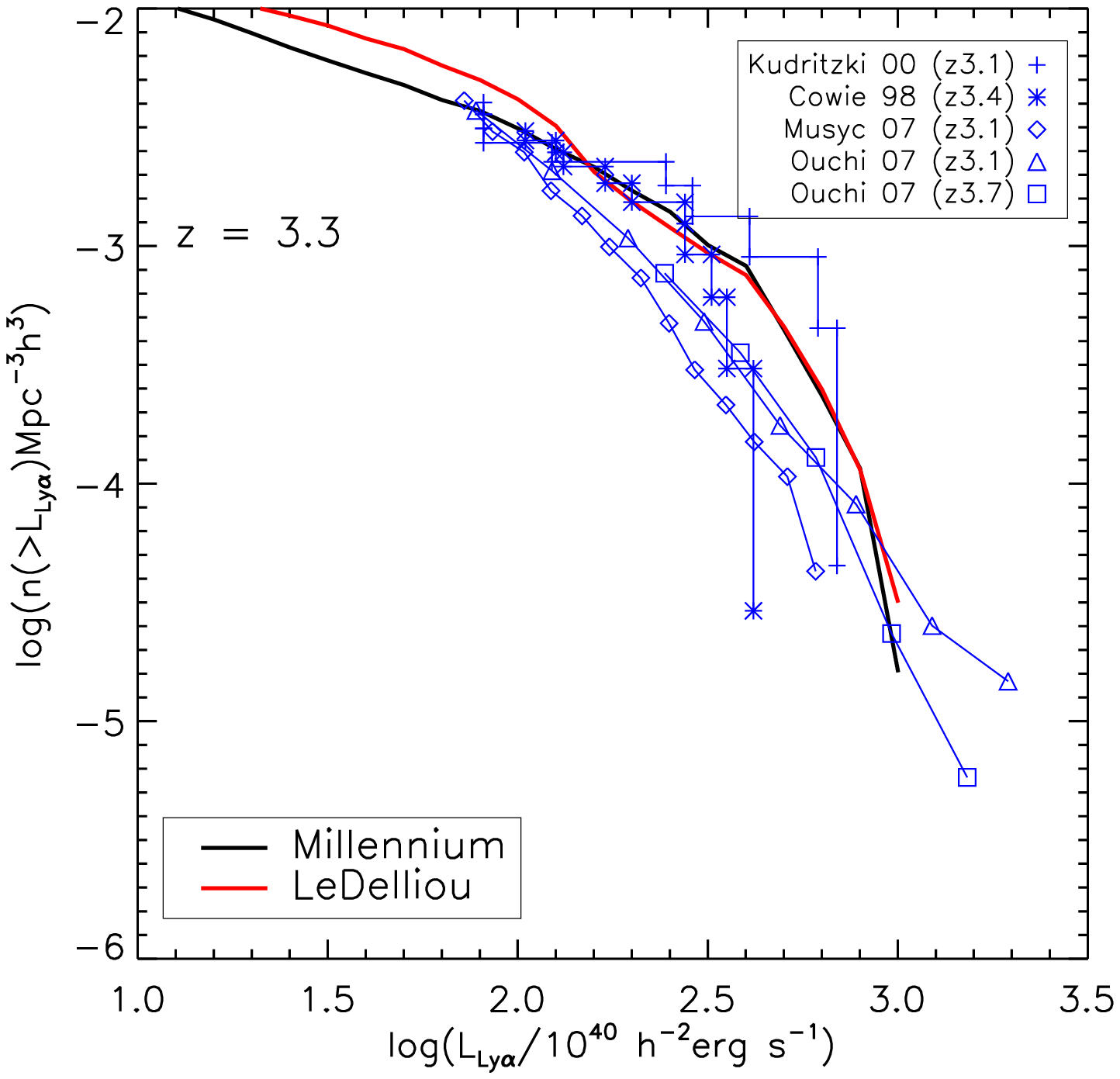}
\includegraphics[width=7cm,angle=90]{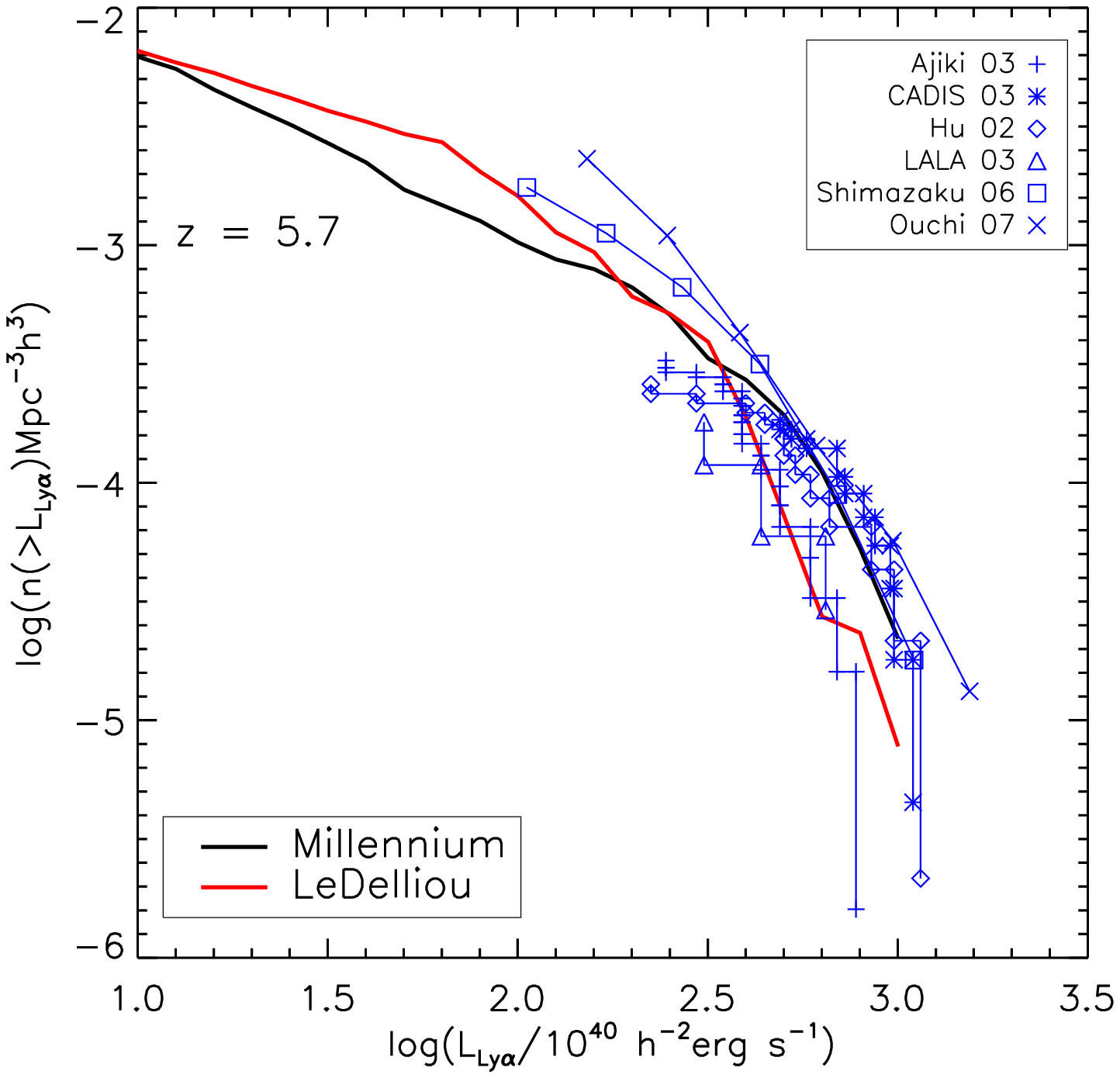}
\includegraphics[width=7cm,angle=90]{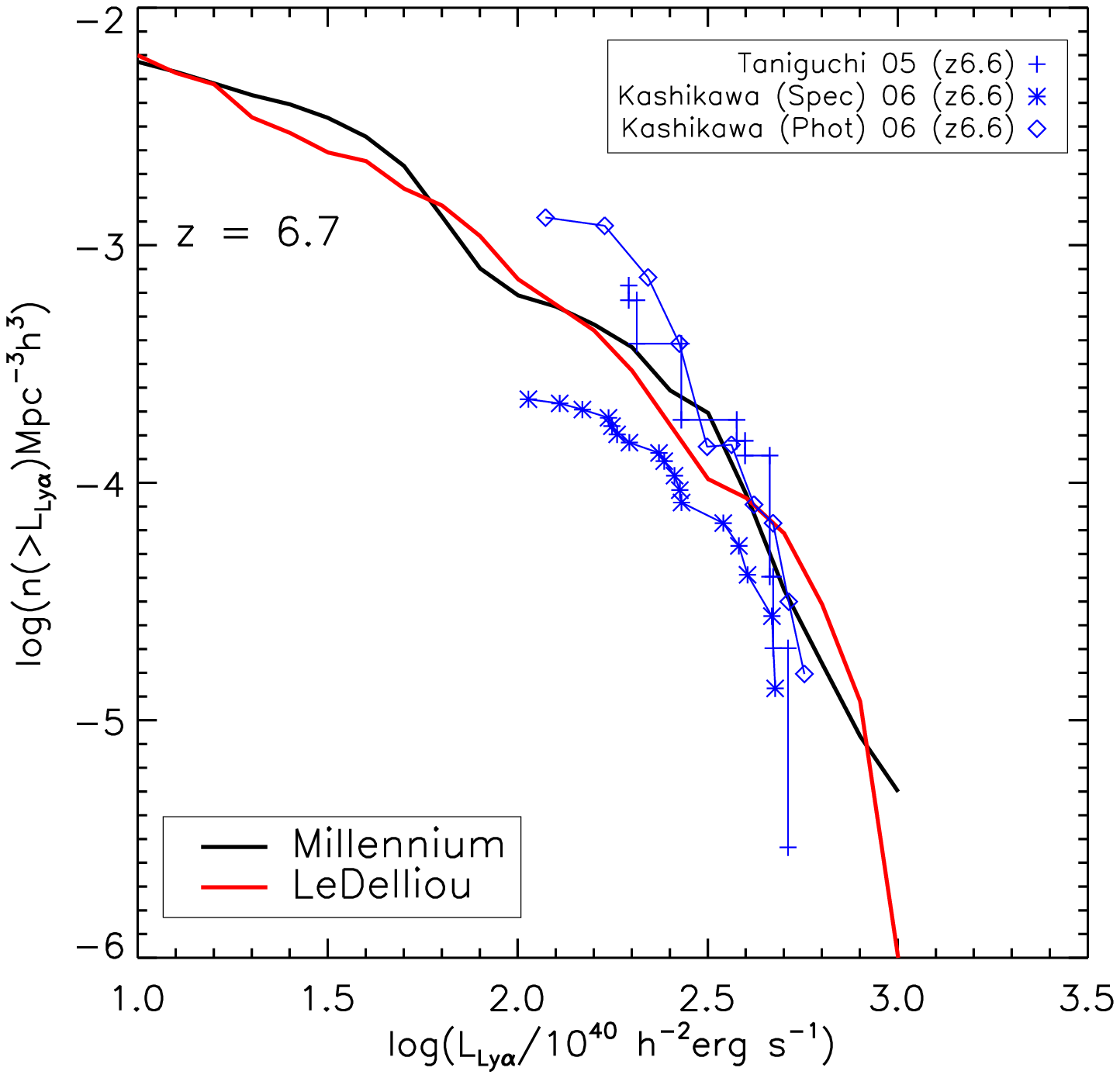}
\caption{The cumulative luminosity functions of \lya\ emitters at 
redshifts $z=3.3$ (Top), $z=5.7$ (Center) and $z=6.7$ (Bottom). 
The blue points correspond to observational data (as indicated by the 
key with full references in the text). The black and red curves correspond, 
respectively, to the \galform\ predictions using the cosmological parameters of 
the Millennium Simulation and those adopted in Le Delliou et al.}
\label{fig.clf}
\end{figure}

In this section, we investigate the impact on the model predictions of
the choice of cosmological parameters by re-running the model of
\citet{dell1, dell2}, keeping the galaxy formation parameters the same
but changing the cosmological parameters to match those used in the
Millennium simulation. To recap, the original \citet{dell2} model used
$\Omega_{\rm{m}} = 0.3$, $\Omega_{\Lambda} = 0.7$, $\Omega_{\rmn{b}} =
0.04$ , $\sigma_8 = 0.93$ and $h = 0.7$.  In Fig.~\ref{fig.clf}, we
compare the cumulative luminosity functions obtained with \galform\
for the two sets of cosmological parameters with current observational
data in the redshift range $3<z<7$. The observational data is taken
from: \citet{kudritzki00} (crosses), \citet{cowie98} (asterisks),
\citet{gawiser07} (diamonds), \citet{ouchi07} (triangles and squares)
in the $z=3.3$ panel; \citet{ajiki03} (pluses), \citet{maier03}
(asterisks), \citet{hu04} (diamonds), \citet{rhoads03} (triangles),
\citet{shimasaku06} (squares) and \citet{ouchi07} (crosses) in the
$z=5.7$ panel; and \citet{taniguchi05} (crosses) and
\citet{kashikawa06} (asterisks and diamonds) in the $z=6.7$ panel. At
$z=3.3$, the two model curves agree very well, and are consistent with
the observational data shown. At $z=5.72$, the two models do not match
as well as in the previous case, but both are still consistent with
the observational data. Finally, at $z=6.7$ the differences are small
and both curves are consistent with observational data.  The
conclusion from Fig.~\ref{fig.clf} is that there is not a significant
change in the model predictions on using these slightly different
values of the cosmological parameters.  Furthermore, the observational
data is not yet sufficiently accurate to distinguish between the two
models or to motivate the introduction of further modifications to
improve the level of agreement, such as using a different \lya\ escape
fraction.

\subsection{The completeness of the Millennium galaxy catalogues}

The Millennium simulation has a halo mass resolution limit of $1.72
\times 10^{10} h^{-1}M_{\odot}$. In a standard \galform\ run, a grid
of haloes which extends to lower mass haloes than the Millennium
resolution is typically used, with $M_{\rm res} = 5 \times 10^{9}
h^{-1} M_{\odot}$ at $z=0$. A fixed dynamic range in halo mass is
adopted in these runs, but with the mass resolution shifting to
smaller masses with increasing redshift: for our standard setup, we
have $M_{\rm res} = 7.8\times 10^7 h^{-1} M_{\odot}$ and $1.4\times
10^7 h^{-1} M_{\odot}$ at $z=3$ and $6$ respectively.  Therefore, when
putting \galform\ galaxies into the Millennium, our sample does not
contain galaxies which formed in haloes with masses below the
resolution limit of the Millennium. This introduces an incompleteness
into our catalogues when compared to the original \galform\
prediction. The incompleteness of the galaxy catalogues is more severe
for low luminosity galaxies because they tend to be hosted by low mass
haloes, as will be shown in the next section. Hereafter, we will use
\textit{N-body sample} to refer to the \galform\ galaxies planted in
the Millennium haloes, to distinguish them from the \textit{pure}
\galform\ catalogues generated using a grid of halos masses.

\begin{figure} 
\centering
\includegraphics[width=7cm]{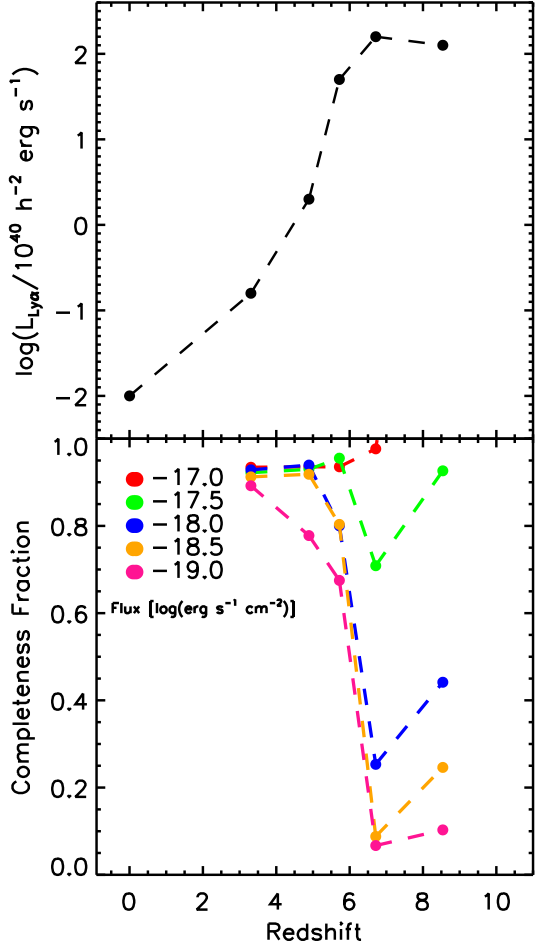}
\caption{Completeness of the Millennium galaxy catalogues with respect
to \lya\ luminosity or flux. (Top): The minimum luminosity down to
which the catalogues are 85$\%$ complete. (Bottom): The completeness
fraction as a function of redshift for a range of fluxes $-19 <
log(\flya \funits) <-17$, as indicated by the key.}
\label{fig.completeness}
\end{figure}

In order to quantify the incompleteness of the N-body sample as a
function of luminosity, we define the completeness fraction as the
ratio of the cumulative luminosity function for the N-body sample to
that obtained for a pure \galform\ calculation, and look for the
luminosity at which the completeness fraction deviates from unity.
The panels of Fig.~\ref{fig.completeness} give different views of the
completeness of the N-body samples. The top panel shows the luminosity
above which a catalogue can be considered as complete: we define the
completeness limit as the luminosity at which the completeness
fraction first drops to $0.85$. The figure clearly shows how the
luminosity corresponding to this completeness limit becomes
progressively brighter as we move to higher redshifts. For $z>9$ the
N-body sample is incomplete at all luminosities plotted.

The bottom panel of Fig.~\ref{fig.completeness} shows how the sample
becomes more incomplete at any redshift as we consider fainter
fluxes. A sample with galaxies brighter than $\log(\flya \funits) =
-19)$ is less than $70\%$ complete at all redshifts $z>5$, while a sample
with galaxies brighter than $\log(\flya \funits) = -17$ is always over
$90\%$ complete for $z<9$. The completeness fraction monotonically
decreases with increasing redshift until $z \sim 6$ for very faint 
fluxes. For $z>6$ the completeness rises again: the shape of the bright
end of the luminosity function at this redshift is sensitive to the choice
of the redshift of reionization.

In summary, the requirement that our samples be at least $80\%$
complete restricts the range of validity of the predictions from the
Millennium simulation to redshifts below $9$, and fluxes brighter than
$\log(\flya \funits)>-17.5$.

\section{Clustering Predictions}

In this section we present clustering predictions using \lya\ emitters in the full 
Millennium volume. To study the clustering of galaxies we calculate the two-point 
correlation function, $\xi(r)$, of the galaxy distribution. In order to quantify the 
evolution of the clustering of galaxies, we measure the correlation function over 
the redshift interval $0<z<9$. 

To calculate $\xi(r)$ in the simulaation, we use the standard
estimator (e.g. Peebles 1980):
\begin{equation}
1+\xi(r) = \frac{\langle DD \rangle }{\frac{1}{2} N_{\rm gal} n \Delta V(r)},
\label{eq.xi1}
\end{equation}
where $\langle DD \rangle$ stands for the number of distinct data
pairs with separations in the range $r$ to $r + \Delta r$, $n$ is the
mean number density of galaxies, $N_{\rm gal}$ is the total number of
galaxies in the simulation volume and $\Delta V(r)$ is the volume of a
spherical shell of radius $r$ and thickness $\Delta r$. This estimator
is applicable in the case of periodic boundary conditions. In the
correlation function analysis, we consider two parameters which help
us to understand the clustering behaviour of \lya\ galaxies: the
correlation length, $r_0$, and the galaxy bias, $b$, both of which are
discussed below.

\subsection{Correlation Length evolution}

A common way to characterize the clustering of galaxies is to fit a power-law 
to the correlation function:
\begin{equation}
\xi(r) = \left( \frac{r}{r_0} \right)^{-\gamma},
\label{eq.xifit}
\end{equation}
where $r_0$ is the correlation length and $\gamma=1.8$ gives a good
fit to the slope of the observed correlation function over a
restricted range of pair separations around $r_0$ at $z=0$
(e.g. \citet{DavisPeebles83}). The correlation length can also be
defined as the scale where $\xi = 1$, and quantifies the amplitude of
the correlation function when the slope $\gamma$ is fixed.

\begin{figure}
\centering
\includegraphics[width=11cm,angle=90]{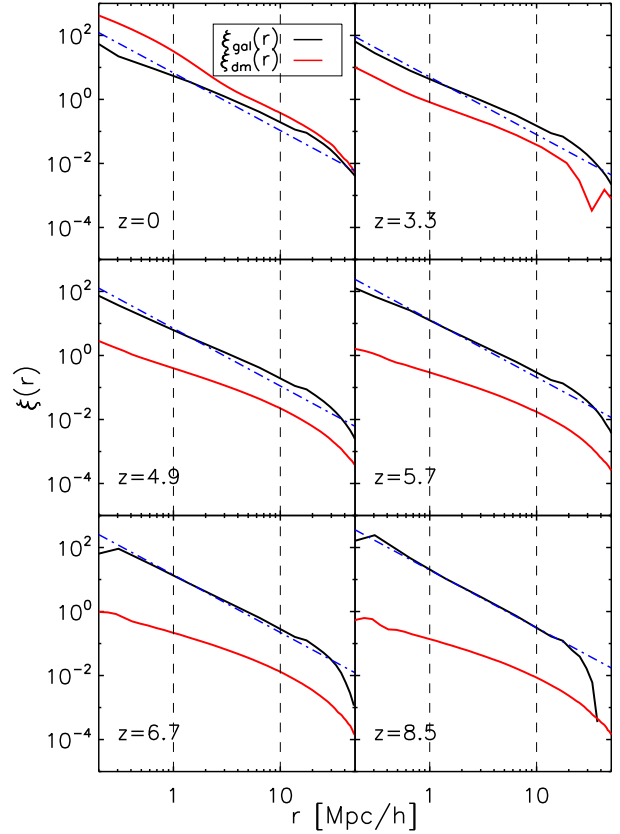}
\caption{The correlation function predicted for \lya\ emitters (black
solid curve) for a range of redshifts, as indicated in each
panel. \lya\ emitters are included down to the completeness limit at
each redshift shown in Fig~\ref{fig.completeness}.  The solid red
curve shows the correlation function of the dark matter at the same
epochs. The blue dashed line shows the power law fit of
Eq.~\eqref{eq.xifit}, evaluated in the range $1<r[\rmn{Mpc/h}]<10$, as
delineated by the vertical dashed lines.}
\label{fig.fullxi}
\end{figure}

Fig.~\ref{fig.fullxi} shows the correlation function of \lya\ emitting
galaxies, \xigal\ (solid black curves) of the full catalogues down to
the completeness limits at each redshift, calculated using
Eq.~\eqref{eq.xi1}. The red curve shows \xidm, the correlation
function of the dark matter. At $z=0$, \xidm\ is larger than \xigal,
but for $z>0$ \xidm\ is increasingly below \xigal.  We will study in
detail the comparison of the dark matter and \lya\ galaxy correlation
functions in \S4.2.

\begin{figure}
\centering
\includegraphics[width=12cm,angle=90]{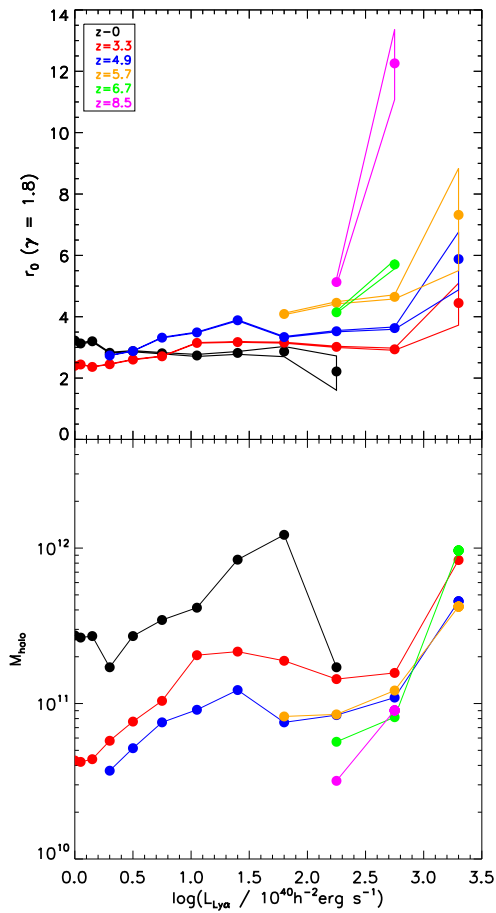}
\caption{\textit{(Top):} The evolution of the correlation length $r_0$
as a function of \lya\ luminosity for several redshifts in the
range $0<z<9$, as indicated by the key. The thin solid coloured lines
shows the errors on the correlation length.  \textit{(Bottom):} The
evolution of the median mass of halos which host \lya\ emitting
galaxies as a function of \lya\ luminosity, for the same range
of redshifts as above.  }
\label{fig.r0_mhalo}
\end{figure}

Another notable feature of Fig.~\ref{fig.fullxi} is that $\xigal(r)$
differs considerably from a power law, particularly on scales greater
than 10 \mpc.  When fitting Eq.~\eqref{eq.xifit} to the correlation
functions plotted in Fig.~\ref{fig.fullxi}, we use only the
measurements in the range [1,10] \mpc, where $\xigal(r)$ behaves most
like a power law. We fix the slope $\gamma = 1.8$ for all $\xigal(r)$
to allow a comparison between different redshifts, although we note
that for $z<5$, the slope of $\xigal(r)$ is closer to $\gamma = 1.6$.
By using the power law fit we can compare the clustering amplitudes of
different galaxy samples. To determine the clustering evolution of
\lya\ emitters, we split the catalogues of \lya\ emitters into
luminosity bins. For each of these sub-samples, we calculate the
correlation function and then we obtain $r_0$ by fitting
Eq.~\eqref{eq.xifit} as described. Fig.~\ref{fig.r0_mhalo} (top) shows
the dependence of $r_0$ on luminosity for different redshifts in the
range $0<z<9$. The errors are shown by the area enclosed by the thin
solid lines for each set of points, and are calculated as the $90\%$
confidence interval of the $\chi^2$ fit of the correlation functions
to Eq.~\eqref{eq.xifit} (ignoring any covariance between pair
separation bins). The range of luminosities plotted is set by the
completeness limit of the simulation described in the previous
section. We also discard galaxy samples with fewer than 500 galaxies,
as in such cases, the errors are extremely large and the correlation
functions are poorly defined. The clustering in high redshift surveys
of \lya\ emitters is sensitive to the flux limit that they are able to
reach, as shown by Fig.~\ref{fig.r0_mhalo}.

The model predictions show modest evolution of $r_0$ with redshift for
most of the luminosity range studied. Over this redshift interval, on
the other hand, the correlation length of the dark matter changes
dramatically, as shown by Fig.~\ref{fig.fullxi}.  Typically, at a
given redshift, we find that $r_0$ shows little dependence on
luminosity until a luminosity of \llya $\sim 10^{42} \lunits$ is
reached, brightwards of which there is a strong increase in clustering
strength with luminosity. This trend is even more pronounced at higher
redshifts. Galaxies at $z=0$ are less clustered than galaxies in the
range $3<z<7$, except at luminosities close to \llya $\sim 10^{40}
\lunits$. At $z=8.5$, $r_0$ increases from $r_0 \sim 5$ \mpc\ at \llya
$\sim 10^{42} \lunits$ to $r_0 \sim 12$ \mpc\ at \llya $> 10^{42.5}
\lunits$.

The growth of $r_0$ with limiting luminosity is related to the masses
of the haloes which host \lya\ galaxies. As shown in the bottom panel
of Fig.~\ref{fig.r0_mhalo}, there is not a simple relation between the
median mass of the host halo and the luminosity of \lya\ emitters.
For a given luminosity, \lya\ galaxies tend to be hosted by haloes of
smaller masses as we go to higher redshifts. In addition, for
redshifts $z>0$, there is a trend of more luminous \lya\ emitters
being found in more massive haloes. The key to explaining the trends
in clustering strength is to compare how the effective mass of the
haloes which host \lya\ emitting galaxies is evolving compared to the
typical or characteristic mass in the halo distribution ($M_{*}$)
\citep{mo96}; if \lya\ emitters tend to be found in haloes more
massive than $M_*$, then they will be more strongly clustered than the
dark matter. This difference between the clustering amplitude of
galaxies and mass is explored more in the next section.  In a
hierarchical model for the growth of structures, haloes more massive
than $M_*$ are more clustered, and thus we expect a strong connection
between the evolution of $r_0$ and the masses of the
halos. Fig.~\ref{fig.r0_mhalo} shows that the dependence of $r_0$ (and
host halo mass) on luminosity becomes stronger at higher redshifts.

\subsection{The bias factor of \lya\ emitters}
\label{sec:bias}

The galaxy bias, $b$, quantifies the strength of the clustering of
galaxies compared to the clustering of the dark matter. One way to
calculate the bias is by taking the ratio of \xigal\ and \xidm,
$\xigal = b^2\xidm$. Both correlation functions are estimated using
Eq.~\eqref{eq.xi1}. Since the simulation contains ten billion dark
matter particles, a direct pair-count calculation of \xidm\ would
demand a prohibitively large amount of computer time, so we extract
dilute samples of the dark matter particles, selecting randomly $\sim
10^7$ particles. In this way we only enlarge the pair-count errors on
\xidm\ (which nevertheless are still much smaller than for \xigal) but
obtain the correct amplitude of correlation function itself.

\begin{figure}
\centering
\includegraphics[width=10cm, angle=90]{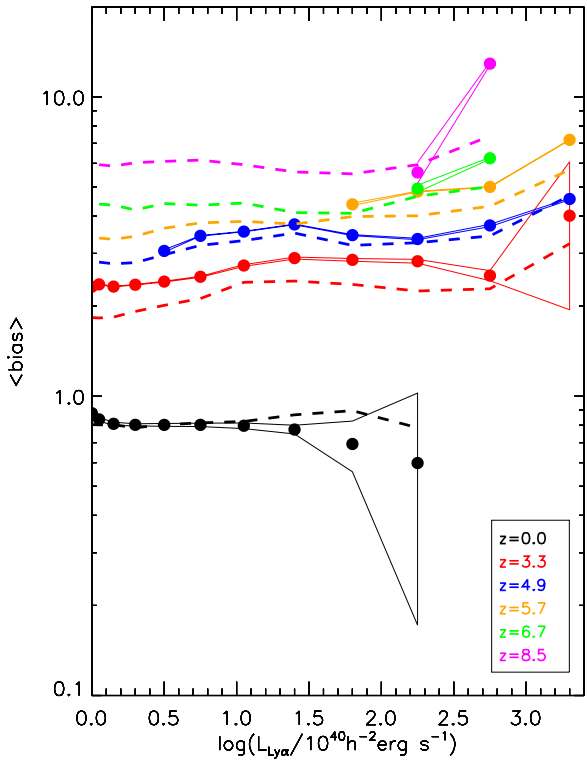}
\caption{The galaxy bias as a function of \lya\ luminosity at
different redshifts, as indicated by the key. The solid lines show the
results from the simulation and the dashed lines show the analytical
expression of SMT. The area enclosed by the thin solid lines shows the
error on the bias estimation for each redshift.}
\label{fig.gal_bias}
\end{figure}

To obtain the bias parameter of \lya\ emitters as a function of
luminosity, we split the full catalogue of galaxies at each redshift
into luminosity bins. For each of these bins we calculate
\xigal\ and divide by \xidm\ to get the sqaure of the bias.  Due to
non-linearities, the ratio of \xigal\ and \xidm\ is not constant on
all scales.  As a reasonable estimation of the bias we chose the mean
value over the range 6 \mpc\ $<r<$ 30 \mpc. Over these scales the bias
does seem to be constant and independent of scale. This range is quite
similar to the one used by \citet{gao05} to measure the bias parameter
of dark matter haloes in the Millennium Simulation.

The bias parameter can also be calculated approximately using various
analytical formalisms \citep{mo96,sheth01,mandelbaum05}. These
procedures relate the halo bias to $\sigma(m,z)$, the \textit{rms}
linear mass fluctuation within a sphere which on average contains mass
$m$. The bias factor for galaxies of a given luminosity is then
obtained by averaging the halo bias over the halos hosting these
galaxies. \citet{dell2} used the analytical expression of
\citet{sheth01} (hereafter SMT) to calculate the bias parameter for
the semi-analytical galaxies. This gives a reasonable approximation to
the large-scale halo bias measured in N-body simulations (e.g
\citet{angulo08}).

Fig.~\ref{fig.gal_bias} shows the bias parameter as a function of
luminosity for redshifts in the range $0<z<9$, and compares the direct
calculation using the N-body simulation (solid lines) with the
analytical estimation (dashed lines).  In order to calculate the
uncertainty in our value of the bias, we assume an error on
$\xigal(r)$ of the form $\Delta \xi_{\rmn{gal}} =
2\sqrt{(1+\xi_{\rmn{gal}})/DD}$ \citep{hewitt82, baugh96}, and
assuming a negligible error in \xidm\ we get
\begin{equation}
 \Delta b = \frac{1}{b\xi_{\rmn{dm}}}\sqrt{\frac{1+b^2\xi_{\rmn{dm}}}{DD}},
\end{equation}
for the error in the bias estimation.  This error is shown in
Fig. \ref{fig.gal_bias} as the range defined by the thin solid lines
surrounding the bias measurement shown by the points.

The first noticeable feature of Fig.~\ref{fig.gal_bias} is the strong
evolution of bias with increasing redshift: From $z=0$ to $z=8.5$ the
bias factor increases from $b(z=0)\sim 0.8$ to $b(z=8.5)\sim 12$,
which means that the clustering amplitude of \lya\ emitters at $z=8.5$
is over 140 times the clustering amplitude of the dark matter at this
redshift.  Another interesting prediction is the dependence of bias on
\lya\ luminosity. For $z > 3$ there seems to be a strong increase of
the bias with luminosity for bins where \llya\ $ > 10^{42}
\lunits$. The agreement between the analytic calculation of the bias
and the simulation result is reasonable over the range $0<z<5$, but
becomes less impressive as higher biases are reached. A similar
discrepancy was also noticed by \citet{gao05}, where they compared the
halo bias extracted from the simulation with different analytic
formulae (see also \citet{angulo08}).

Another way to describe galaxy clustering is through the halo
occupation distribution (HOD; \citet{benson00}, \citet{berlind03},
\citet{cooray02}). The HOD gives the mean number of galaxies which
meet a particular observational selection as a function of halo
mass. For flux-limited samples, the HOD can be broken down into the
contribution from central galaxies and satellite galaxies. In a simple
picture, the mean number of central galaxies is zero below some
threshold halo mass, $M_{min}$, and unity for higher halo masses. With
increasing halo mass, a second threshold is reached, $M_{crit}$, above
which a halo can also host a satellite galaxy. The number of
satellites is usually described by a power-law of slope $\beta$. In
the simplest case, three parameters are needed to describe the HOD
\citep{berlind02,hamana04}; more detailed models have been proposed to
describe the transition from 0 to 1 galaxy \citep{berlind03}.

\begin{figure}
\centering
\includegraphics[width=12cm,angle=90]{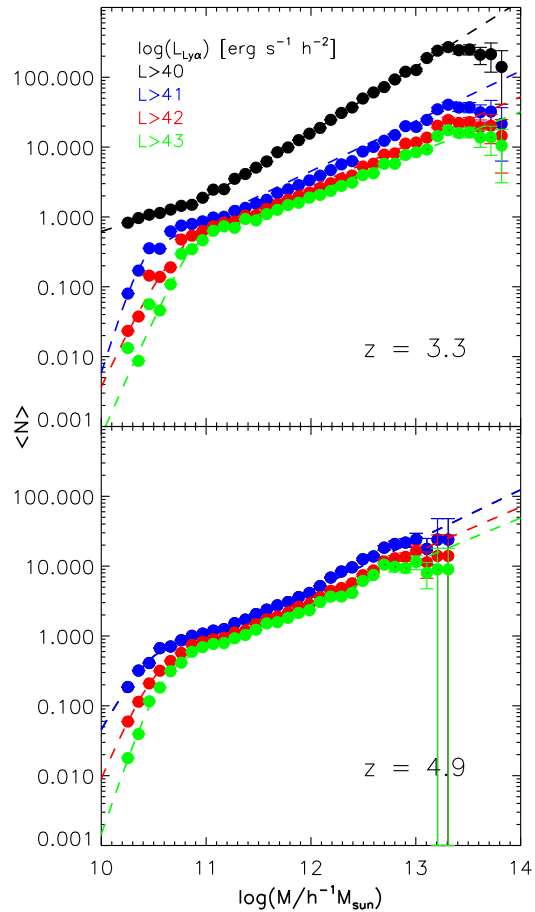}
\caption{The HOD of \lya\ emitters at $z= 3.3$ ({\it top}) and $z =
4.9$ ({\it bottom}).  Each set of points represents a model sample
with a different luminosity limit, as given by the key in the upper
panel. The dashed line in each panel correspond to a ``best'' fit
using the Berlind et al. (2003) parametrization. }
\label{fig.hod}
\end{figure}

We can compute the HOD directly from our model. The results are shown
in Fig.~\ref{fig.hod}, where we plot the HOD at two different
redshifts for different luminosity limits. For comparison, we plot the
HOD parametrization of \citet{berlind03} against our model
predictions.  In general, this HOD does a reasonable job of describing
the model output, and is certainly preferred over a simple three
parameter model. However, for the $z=3.3$ case (top panel of
Fig.~\ref{fig.hod}), the shape of the model HOD for $\log
(M/M_{\odot}) > 13$ is still more complicated than can be accommodated
by the Berlind et al. parametrization, showing a flattening in the
number of satellites as a function of increasing halo mass.  There is
less disagreement in the $z=4.9$ case (bottom panel), but our model
HOD becomes very noisy for large halo masses.

\section{Mock Catalogues}

\begin{table*}
\caption{Summary of survey properties and simulation results.}
\label{table.mock}
\begin{tabular}{@{}lcccccccccc}
\hline
\hline
 (1)  & (2)  & (3)   &(4)  &(5) &(6)              &(7)    &(8)  &(9) &(10) &(11)\\
\hline
Survey & $z_{\rmn{survey}}$ & $z_{\rmn{simulation}}$ & $\Delta z$ & Area $\arcm^2$ & 
\ewobs [\AA{}] & \flya \funits  & $N_{\rmn{obs}}$ & $N^{\rmn{median}}_{\rm{mock}} $  & 10-90\% & $C_v$\\
\hline
MUSYC & 3.1 & 3.06 & 0.04 & 961 & 80 & $1.5 \times 10^{-17}$ & 162 & 142 & 89-207
& 0.41\\
SXDS  & 3.1 & 3.06 & 0.06 & 3538 & 328 & $1.1 \times 10^{-17} $ & 356 & 316 &
256-379 & 0.19\\
      & 3.7  & 3.58 & 0.06 & 3474 & 282 & $2.7 \times 10^{-17}$ & 101 & 80 &
60-110 & 0.31\\
      & 5.7  & 5.72 & 0.10 & 3722 & 335 & $7.4 \times 10^{-18}$ & 401 & 329 &
255-407 & 0.23\\
ELVIS & 8.8  & 8.54 & 0.10 & $\sim 3160$ &100& $3.7 \times 10^{-18} $ & -- & 20 &
14-29 & 0.37\\
\hline
\hline
\end{tabular}
\\
\begin{flushleft}
\indent Column (1) gives the name of the survey; (2) and (3) show the
redshift of the observations and nearest output from the simulations,
respectively; (4) shows the redshift width of the survey, based on the
FWHM filter width; (5) shows the area covered by each survey; (6) and
(7) show the equivalent width and \lya\ flux limits of the samples,
respectively; (8) shows the number of galaxies detected in each
survey; (9) and (10) show the median of the number of galaxies and the
10-90 percentile range found in the mock catalogues for each
survey. Finally, column (11) gives the fractional variation of the
number of galaxies, defined in Eq.~\eqref{eq.cv}.
\end{flushleft}
\end{table*}

In this section we make mock catalogues of \lya\ emitters for a
selection of surveys. In the previous section, we used the full
simulation box to make clustering predictions, exploiting the periodic
boundary conditions of the computational volume. The simulation is so
large that it can accommodate many volumes equivalent to those sampled
by current \lya\ surveys, allowing us to examine the fluctuations in
the number of emitters and their clustering.  The characteristics of
the surveys we replicate are listed in Table~\ref{table.mock}.

The procedure to build the mock catalogues is the following:

\begin{enumerate}
\item We extract a catalogue of galaxies from an output of the
Millennium Simulation that matches (as closely as possible) the
redshift of a given survey. The simulation output contains 64
snapshots spaced roughly logarithmically in the redshift range
$[127,0]$.
\item We choose one of the axes (say, the z-axis) as the
line-of-sight, and we convert it to \textit{redshift space}, to match
what is observed in real surveys. To do this we replace $r_z$ (the
comoving space coordinate) with
\begin{equation}
 s_z = r_z + \frac{v_z}{a\rm{H(z)}} \quad [\mpc],
\end{equation}
where $v_z$ is the peculiar velocity along the z-axis, $a = 1/(1+z)$
and H($z$) is the Hubble parameter at redshift $z$.
\item We then apply the flux limit of the particular survey, to mimic the selection of 
galaxies. Table \ref{table.mock} shows the flux limits of the surveys considered.
\item Then we extract many mock catalogues using the same geometry as the real survey. 
We extract slices of a particular depth $\Delta z$ (different for each survey), and 
within each slice we extract as many mock catalogues as possible using the same angular 
geometry as the real sample. $\Delta z$ is determined using the transmission curves 
of the narrow-band filters used in each survey. To derive the angular sizes we use:
\begin{equation}
\label{eq.tran_size}
D_t(\theta,z) = d_c(z)  \Delta \theta,
\end{equation}
\begin{equation}
\label{eq.dc}
 d_c(z) = \frac{c}{H_0} \int_0^z \frac{{\rm d}z'}{\sqrt{\Omega_m (1+z')^3 + \Omega_{\Lambda}}},
\end{equation}
where $D_t$ is the transverse comoving size in $h^{-1}$Mpc, $d_c$ is
the comoving radial distance, $c$ and $H_0$ are the speed of light and
the Hubble constant respectively, $\Omega_{\rm m}$ and
$\Omega_{\Lambda}$ are the density parameters of matter and the
cosmological constant respectively. Eq.~\eqref{eq.tran_size} is valid
for $\Delta \theta \ll 1 [\rm{radians}]$, which is the case for the
surveys we analyse in this work. We assume a flat cosmology.
\item From the line-of-sight axis we invert Eq. \eqref{eq.dc} to
obtain the redshift distribution of \lya\ galaxies within each mock
catalogue, converting galaxy position to redshift. This information is
then used to take into account the shape of the filter transmission
curve for each survey, which controls the minimum flux and equivalent
width as a function of redshift. The value given in
Table~\ref{table.mock} corresponds to the minimum flux and \ewobs\ at
the peak of the filter transmission curve.  For redshifts at which the
transmission is smaller (the tails of the curve) the minimum flux and
\ewobs\ required for a \lya\ emitter to be included are proportionally
bigger.
\item Finally, we allow for incompleteness in the detection of \lya\
emitters at a given flux due to noise in the observed images (where
this information is available). To do this, we randomly select a
fraction of galaxies in a given \lya\ flux bin to match the
completeness fraction reported for the survey at that flux.

\end{enumerate}

Real surveys of \lya\ emitters usually lack detailed information about
the position of galaxies along the line-of-sight. Hence, instead of
measuring the spatial correlation function defined in
Eq.~\eqref{eq.xi1}, it is only possible to estimate the angular
correlation function, \wtheta, which is the projection on the sky of
$\xi(r)$.

We estimate \wtheta\ from mock catalogues using the following
procedure, which closely matches that used in real surveys. To compute
the angular correlation function we use the estimator \citep{landy93}:
\begin{equation}
w_{LS}(\theta) = \frac{\langle DD(\theta) \rangle - 2 \langle DR(\theta) \rangle + \langle RR(\theta) \rangle}{\langle RR(\theta) \rangle},
\label{eq.ls}
\end{equation}
where $\langle DR \rangle$ stands for data-random pairs, $\langle RR
\rangle$ indicates the number of random-random pairs and all of the
pair counts have been appropriately normalized.  In the case of a
finite volume survey, this estimator is more robust than the one
defined in Eq.~\eqref{eq.xi1} because it is less sensitive to errors
in the mean density of galaxies, such as could arise from boundary
effects. In practice, the measured angular correlation function can be
approximated by a power law:
\begin{equation}
w(\theta) = A_w\left(\frac{\theta}{1^{\circ}}\right)^{-\delta},
\label{eq.powerlaw}
\end{equation}
where $A_w$ is the dimensionless amplitude of the correlation
function, and $\delta$ is related to slope of the spatial correlation
function, $\gamma$, from Eq.~\eqref{eq.xifit} by $\delta = \gamma -
1$. A relation between $r_0$ and $A_w$ can be obtained using a
generalization of Limber's equation \citep{simon07}.

Surveys of \lya\ emitters typically cover relatively small areas of
sky and can display significant clustering even on the scale of the
survey. As a result, the mean galaxy number density within the
survey area will typically differ from the cosmic mean value.  If the
number of galaxies within the survey is used to estimate the mean
density, used in Eq.~\eqref{eq.ls}, rather than the unknown true
underlying density, this leads to a bias in the estimated correlation
function. This effect is known as the integral constraint (IC)
bias. \citet{landy93} show that when their estimator is used, the
expected value of the estimated correlation function $w_{LS}(\theta)$
is related to the true correlation function $w(\theta)$ by
\begin{equation}
\langle w_{LS}(\theta) \rangle =  \frac{w(\theta)-w_{\Omega}}{1 +
  w_{\Omega}} ,
\label{eq.xi_ic}
\end{equation}
where the integral constraint term $w_{\Omega}$ is defined as
\begin{equation}
w_{\Omega} \equiv \frac{1}{\Omega^2}\int d\Omega_1 d\Omega_2 w(\theta_{12}), 
\label{eq.ic}
\end{equation}
integrating over the survey area, and is equal to the fractional
variance in number density over that area.

When the clustering is weak Eq. \eqref{eq.xi_ic} simplifies to $\langle
w_{LS}(\theta) \rangle \simeq w(\theta) - w_{\Omega}$. This motivates
the additive IC correction which is customarily used in practice:
\begin{equation}
w_{corr}(\theta) = w_{LS}(\theta) + w_{\Omega}.
\label{eq.wcorr}
\end{equation}
We use this to correct the angular correlation functions from our mock
 catalogues.  In order to estimate the term $w_{\Omega}$, we
 approximate the true correlation function as a power law, as in
 Eq. \eqref{eq.powerlaw}, and use
\begin{equation}
w_{\Omega} \simeq A_w \frac{\sum_i \langle RR_i \rangle \theta_i^{-\delta}}
{\sum \langle RR_i \rangle},
\label{eq.sigma}
\end{equation}
\citep{daddi00}, where $\langle RR \rangle$ are the same random pairs
as used in the estimate of $w_{LS}(\theta)$.

To quantify the sample variance expected for a particular survey, we
use the mock catalogues to calculate a \textit{coefficient of
variance} ($C_v$), which is a measure of the fractional variation in
the number of galaxies found in the mocks
\begin{equation}
 C_v = \frac{N_{90} - N_{10}}{2N_{\rm{med}}},
\label{eq.cv}
\end{equation}
where $N_{10}$ and $N_{90}$ are the 10 and 90 percentiles of the
distribution of the number of galaxies in the mocks, respectively, and
$N_{\rm{med}}$ is the median. The value of $C_v$ allows us to compare
the sampling variance between different surveys in a quantitative way.

To analyse the clustering in the mock catalogues, we measured the
angular correlation function of each mock catalogue using the
procedure explained above.  Then we fit Eq.~\eqref{eq.powerlaw} to
each of the mock \wtheta\ and we choose the median value of $A_w$ as
the representative power law fit. We fix the slope of \wtheta\ to
$\delta = 0.8$ for all surveys, except for ELVIS, where we found that
a steeper slope, $\delta = 1.2$, agreed much better with the simulated
data. To express the variation in the correlation function amplitude
found in the mocks, we calculate the 10 and 90 percentiles of the
distribution of $A_w$ for each set of mock surveys. We also calculate
\wtheta\ using the full transverse extent of the simulation, with the
same selection of galaxies as for the real survey. This estimate of
\wtheta, which we call the \textit{Model} \wtheta, represents an ideal
measurement of the correlation function without boundary effects (so
there is no need for the integral constraint correction).

The surveys we mimic are the following: the MUSYC Survey
\citep{gronwall07, gawiser07}, which is a large sample of \lya\
emitting galaxies at $z=3.1$; the SXDS Survey \citep{ouchi05,ouchi07},
which covers three redshifts: $z=3.1$, $z=3.7$ and $z=5.7$, and
finally, we make predictions for the forthcoming ELVIS survey
\citep{kim07a,kim07b}, which is designed to find \lya\ emitting
galaxies at $z=8.8$.  We now describe the properties of the mock
catalogues for each of these surveys in turn.

\subsection{The MUSYC Survey}

\begin{figure}
\centering
\includegraphics[width=7cm]{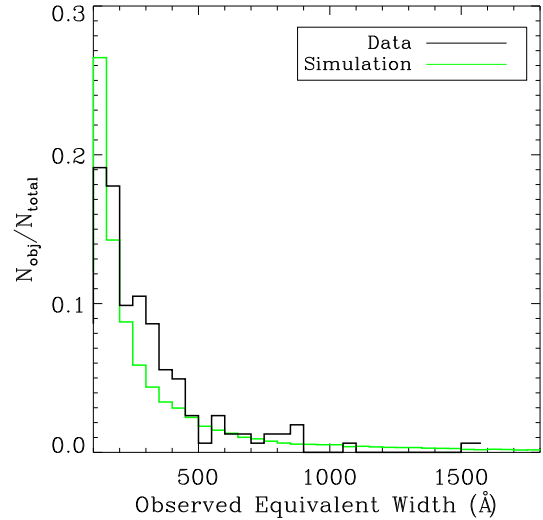}
\caption{The observed $EW_{obs}$ distribution of the MUSYC survey at
$z=3.1$ (solid black line) and the simulation (solid green line).}
\label{fig.ew_musyc}
\end{figure}

\begin{figure}
\centering
\includegraphics[width=8cm]{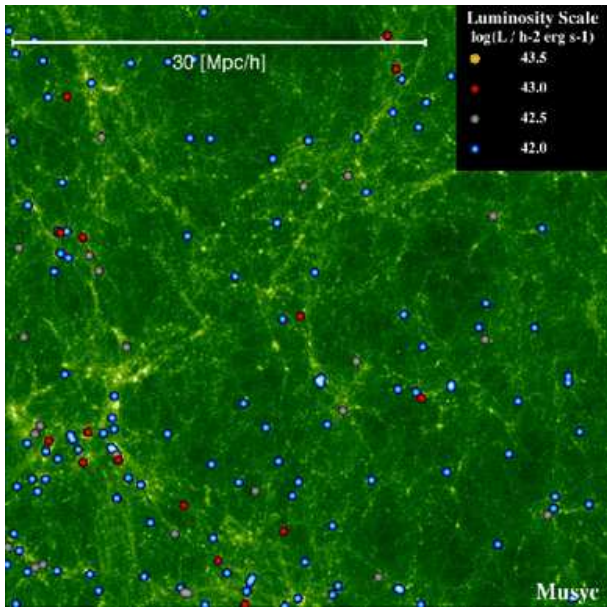}
\caption{An image of a mock catalogue of the MUSYC Survey of \lya\
  emitters at $z=3.1$.  The colour format and legend are the same as
  used in Fig.~\ref{fig.sample}.  The angular size of the image is
  $31\arcmin \times 31 \arcmin$.}
\label{im.musyc}
\end{figure}

The Multi-wavelength Survey by Yale-Chile (MUSYC) \citep{quadri07,
gawiser06,gawiser07,gronwall07} is composed of four fields covering a total solid
angle of one square degree, each one imaged from the ground in the
optical and near-infrared. Here we use data from a single MUSYC field
consisting of narrow-band observations of \lya\ emitters made with the
CTIO 4-m telescope in the Extended Chandra Deep Field South (ECDFS)
\citep{gronwall07}. The MUSYC field, centred on redshift $z=3.1$,
contains 162 \lya\ emitters in a redshift range of $\Delta z \sim
0.04$ over a rectangular area of $31 \arcmin \times 31 \arcmin$ with
flux and \ewobs\ limits described in Table~\ref{table.mock}.

\begin{figure}
\centering
\includegraphics[width=7cm]{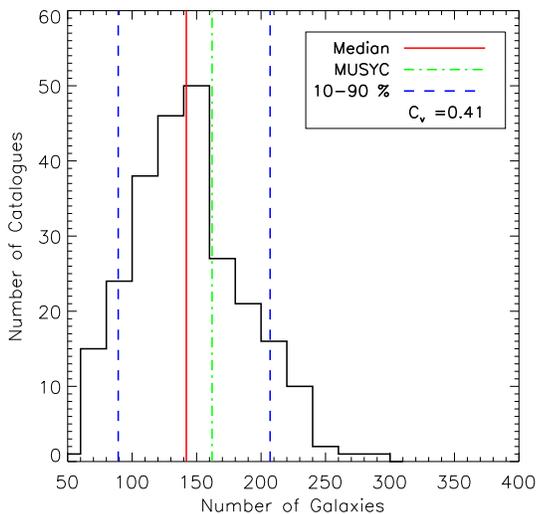}
\caption{Histogram of the number of \lya\ emitters found in the mock
MUSYC catalogues.  The red line shows the median, the dashed blue
lines show the 10-90 percentile range, and the green line shows the
number of galaxies detected in the real survey.}
\label{fig.ngal_musyc}
\end{figure}

To test how well the model reproduces the \lya\ emitters seen in the
MUSYC survey, we first compare the predicted (green) and measured
(black) distributions of \lya\ equivalent widths in
Fig.~\ref{fig.ew_musyc}. Here the predicted distribution comes from
the full simulation volume. Overall, the simulation shows remarkably
good agreement with the real data, with a slight underestimation in
the range $200 < \ewobs [$\AA{}$] < 400$. For $\ewobs[$\AA{}$] > 400$
both distributions seem to agree well, although the number of detected
\lya\ emitters in the tail of the distribution is small.

For the MUSYC survey we built $252$ mock catalogues from the
Millennium simulation volume using the procedure outlined above.
Fig.~\ref{im.musyc} shows an example of one of these mock
catalogues. Many of the \lya\ emitters are found in high dark matter
density regions, and thus they are biased tracers of the dark
matter. Fig.~\ref{fig.ngal_musyc} shows the distribution of the number
of galaxies in the ensemble of mocks. The green line shows the number
detected in the real survey (162), which falls within the 10-90
percentile range of the mock distribution and is close to the median
(142). The 10-90 percentile range spans an interval of
$89<N_{\rm{gal}}<207$, indicating a large cosmic variance for this
survey configuration, with $C_v = 0.41$.

\begin{figure}
\centering
\includegraphics[width=8cm]{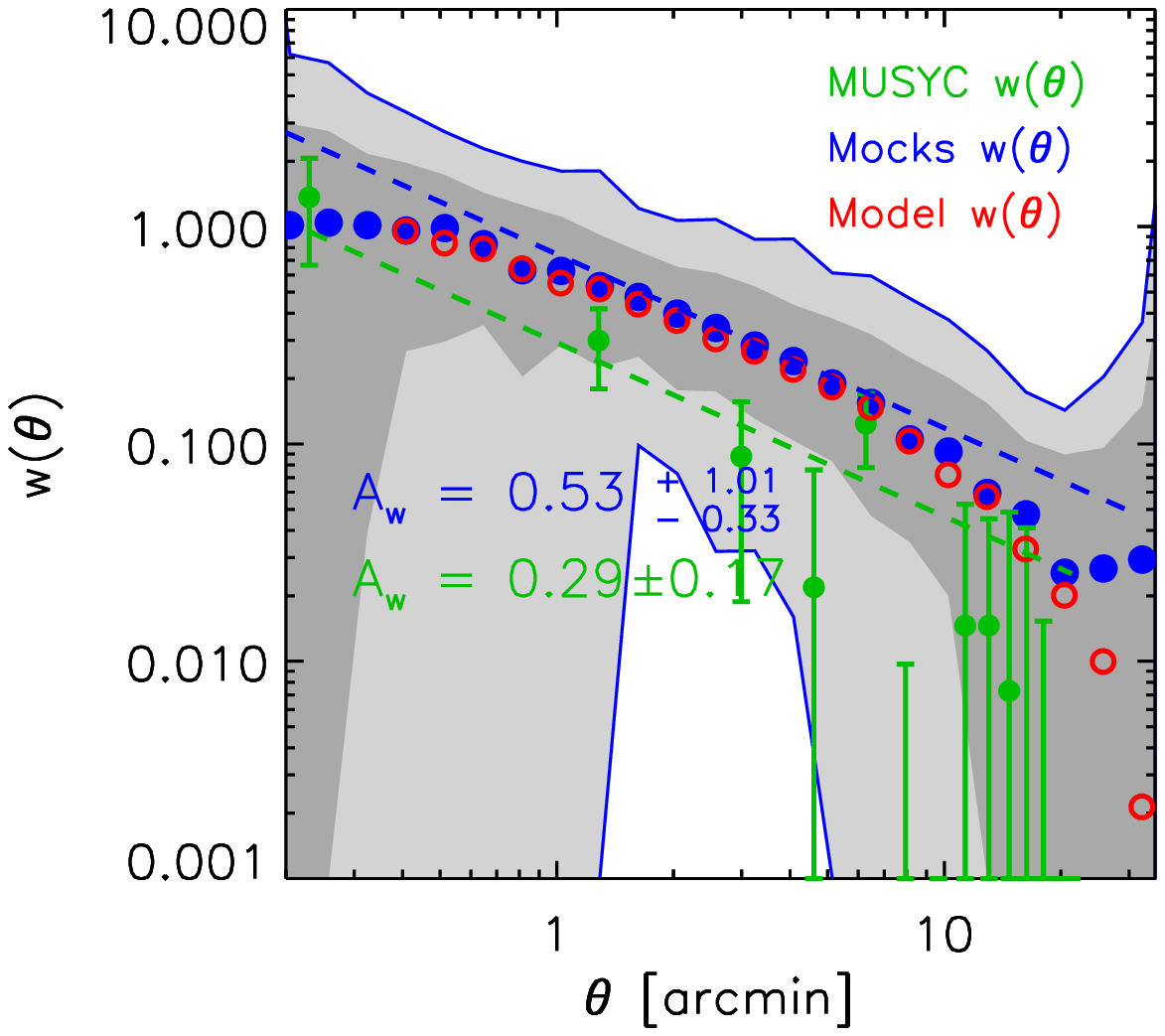}
\caption{ Angular clustering for the MUSYC Survey. Green circles show
\wtheta\ calculated from the observed catalogue \citep{gawiser07}. The
blue circles show the median \wtheta\ from all mock catalogues,
corrected for the integral constraint effect.  The dark and light grey
shaded regions respectively show the 68\% and 95\% ranges of the
distribution of \wtheta\ measured in the mock catalogues. The red open
circles show the \textit{Model} correlation function, obtained using
the width of the entire simulation box (and the same EW, flux and
redshift limits). The dashed lines show the power-law fit to the
observed \wtheta\ (green) and the median fit to \wtheta\ from the mock
catalogues (blue). The amplitudes $A_w$ of these fits are also given
in the figure.}
\label{fig.w_musyc}
\end{figure}

\begin{figure}
\centering
\includegraphics[width=7.2cm]{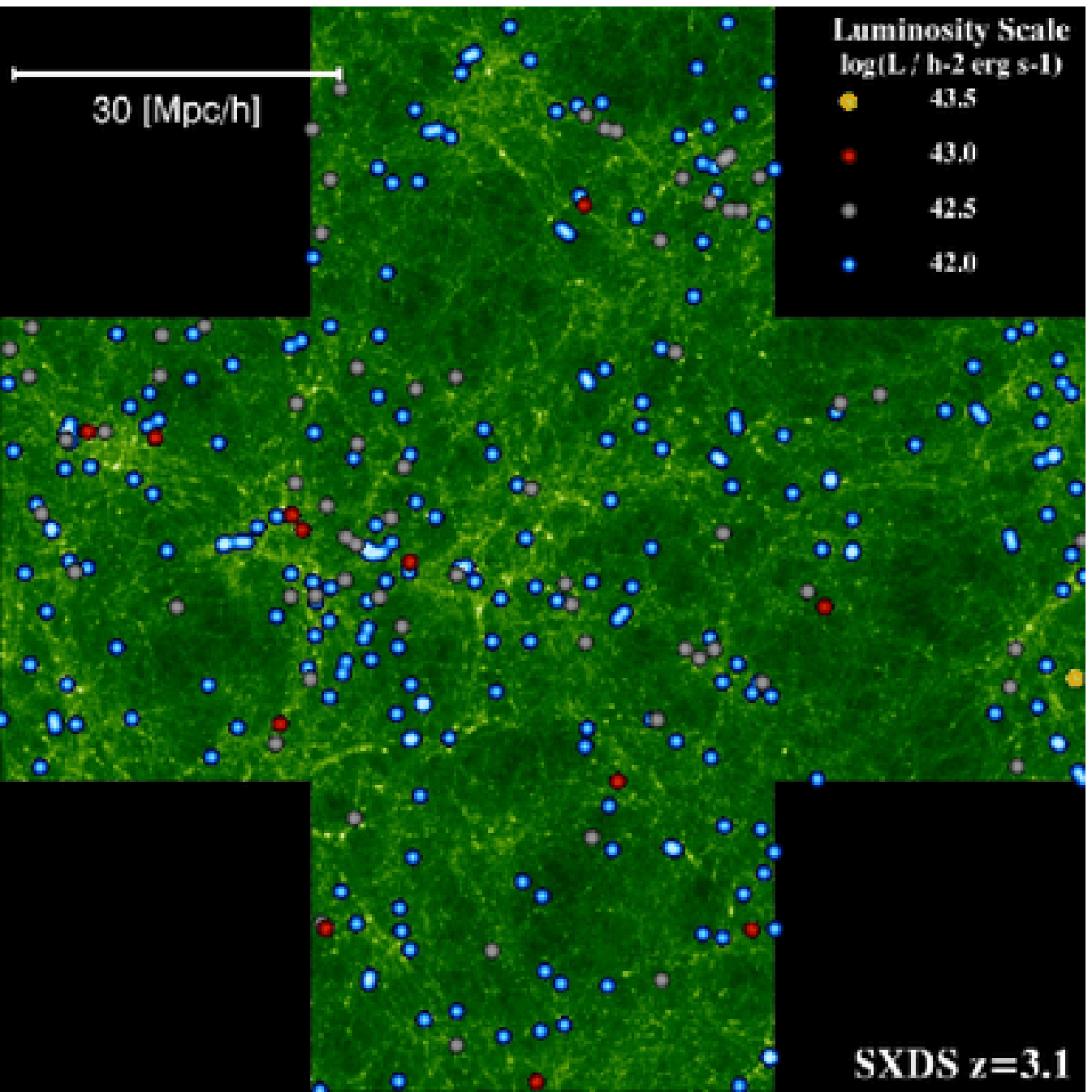}
\includegraphics[width=7.2cm]{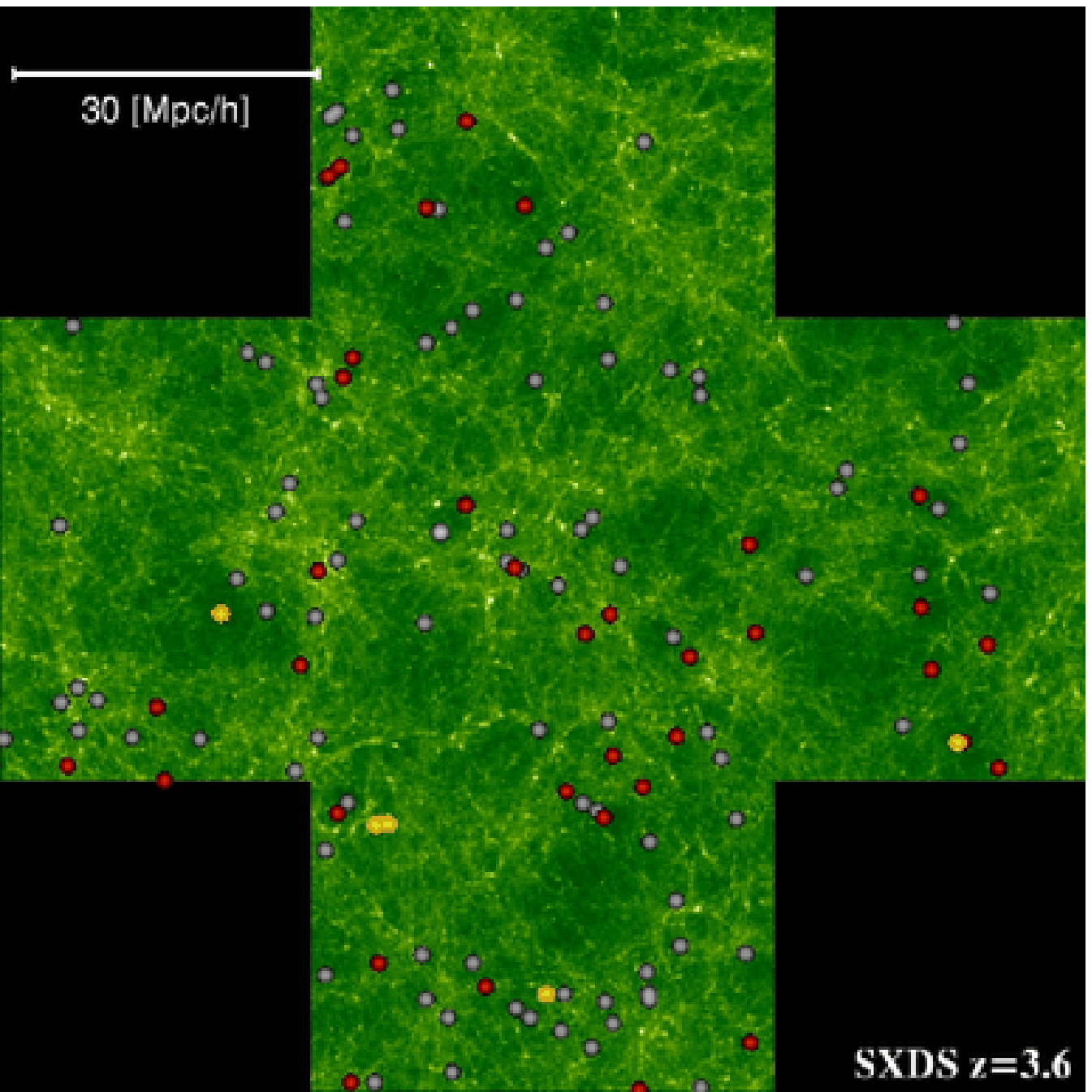}
\includegraphics[width=7.2cm]{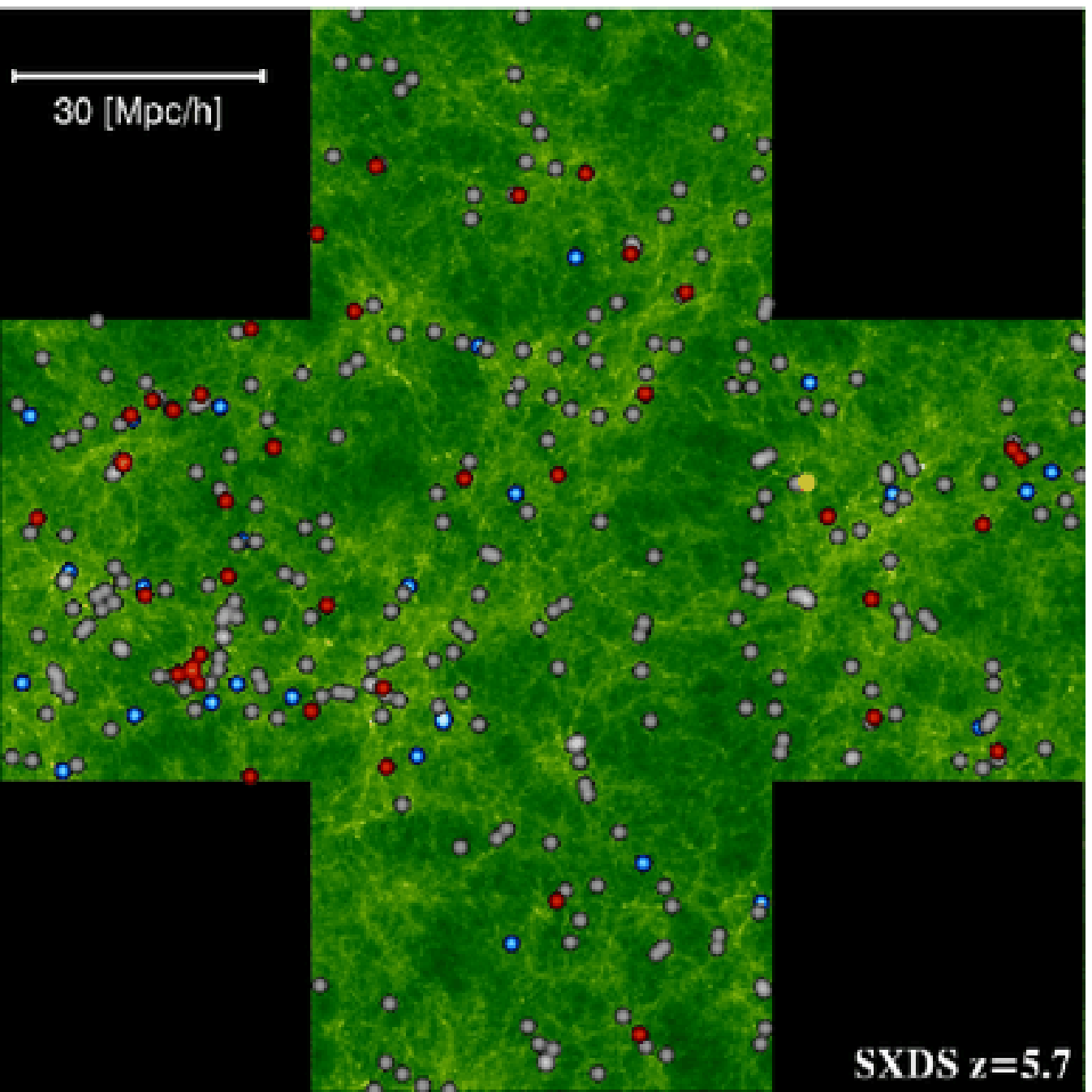}
\caption{Mock catalogues of the SXDS survey for redshifts $3.1$
(\textit{top}), $3.6$ (\textit{centre}) and $5.7$
(\textit{bottom}). The colour scheme and legend are the same as used
previously. The angular size of the image is $1.4^\circ \times
1.4^\circ$.  }
\label{im.sxds}
\end{figure}

The next step is to compare the clustering in the simulations with the
real data.  Fig.~\ref{fig.w_musyc} plots the correlation functions
from the mock catalogues alongside that measured in the real survey
\citep{gawiser07}. There is reasonable agreement between the mock
catalogue results and the observed data \footnotemark.
\footnotetext{After this paper was accepted for publication, we learned that the 
MUSYC data shown in Fig.\ref{fig.w_musyc} are not corrected for the
IC. Including this correction would improve the agreement with the model.}
The median \wtheta\ from the
mocks is slightly higher than the observed values, but the observed
\wtheta\ is within the range containing 95\% of the mock \wtheta\
values (i.e. between the 2.5\% and 97.5\% percentiles, shown by the
light grey shaded region). We quantified this difference by fitting
the power law of Eq.~\eqref{eq.powerlaw} to both real and mock
data. The power-law fits were made over the angular range 1-10
arcmins. We find the value of $A_w$ (Eq.~\ref{eq.powerlaw}) for each
of the mock catalogue correlation functions by $\chi^2$-fitting (using
the same expression as in \S\ref{sec:bias} for the error on each model
datapoint) and then we plot the power law corresponding to the median
value of $A_w$. We find $A_w = 0.53^{+1.01}_{-0.33}$ for the mocks,
where the central value is the median, and the range between the error
bars contains 95\% of the values from the mocks.  For the real data,
we find the best-fit $A_w$ and the 95\% confidence interval around it
by $\chi^2$-fitting, using the error bars on the individual datapoints
reported by \citeauthor{gawiser07}. This gives $A_w = 0.29 \pm 0.17$
for the real data. We again see that the observed value is within the
95\% range of the mocks, and is thus statistically consistent with the
model prediction. We also see that the 95\% confidence error bar on
the observed $A_w$ is much smaller than the error bar we find from our
mocks. This latter discrepancy arises from the small errors quoted on
\wtheta\ by \citet{gawiser07}, which are based on { modified} Poisson pair
count errors, but neglect variations between different sample volumes
(i.e. cosmic variance). On the contrary, using our mocks, we are
able to take cosmic variance fully into account. This underlines the
importance of including the cosmic variance in the error bars on
observational data, to avoid rejecting models by mistake.

The red open circles in Fig.~\ref{fig.w_musyc} show the correlation
function obtained using the full angular size attainable with the
Millennium simulation but keeping the same flux, EW and redshift
limits as in the MUSYC survey (averaging 7 different slices), and so
this measurement has a smaller sample variance. The area used here is
$\sim 120$ times bigger than the MUSYC area, so IC effects are
negligible on the scales studied here. We refer to this as the
\textit{Model} prediction for \wtheta.

The median of the mock correlation functions (including the IC
correction, blue circles) is seen to agree reasonably well with the
\textit{Model} correlation function (red open circles) for $\theta <
20 \arcm$. This shows that for this survey it is possible to obtain an
observational estimate of the correlation function which is unbiased
over a range of scales, by applying the integral constraint
correction. However, on large scales the median \wtheta\ of the mocks
(with IC correction included) lies above the Model \wtheta, which
shows that the IC correction is not perfect, even on
average. Presumably this failure is due (at least in part) to the fact
that the IC correction procedure assumes that \wtheta is a power law,
while the true \wtheta\ departs from a power law on large scales. It
is also important to note that these statements only apply to the
median \wtheta\ derived from the mock samples - the individual mocks
show a large scatter around the true \wtheta\ (as shown by the grey
shading), and the IC correction does not remove this. This scatter
rapidly increases at both small and large angular scales, so the best
constraints on \wtheta\ from this survey are for intermediate scales,
$1 \la \theta \la 5 \arcm$.

\subsection{The SXDS Surveys}

The Subaru/\textit{XMM-Newton} Deep Survey (SXDS)
\citep{ouchi05,ouchi07,kashikawa06} is a multi-wavelength survey
covering $\sim 1.3$ square degrees of the sky. The survey is a
combination of deep, wide area imaging in the X-ray with
\textit{XMM-Newton} and in the optical with the Subaru
Suprime-Cam. Here we are interested in the narrow-band observations at
three different redshifts: $3.1, 3.6$ and $5.7$ \citep{ouchi07}.

\begin{figure*}
\centering
\includegraphics[width=6.3cm,angle=90]{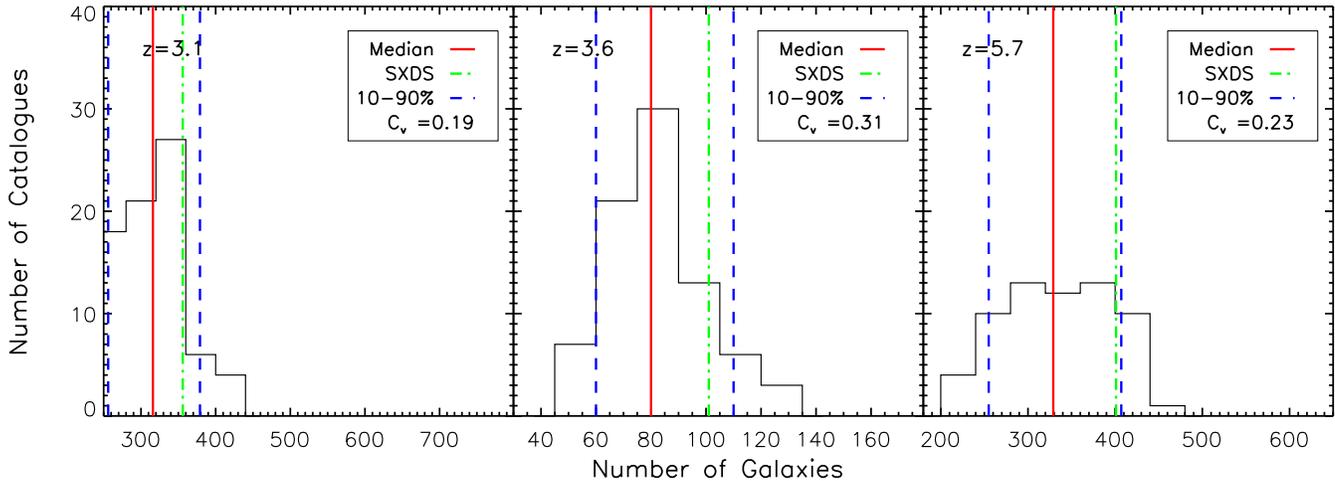}
\caption{The distribution of the number of galaxies in mock SXDS
catalogues, for $z=3.1$ \textit{(left)}, $z=3.6$ \textit{(centre)} and
$z=5.7$ \textit{(right)}. The red line shows the median of the number of
galaxies inside the mock catalogues, the blue lines show the 10-90
percentiles of the distribution, and the green line shows the
number observed in the SXDS.}
\label{fig.ng_sxds}
\end{figure*}

\begin{figure*}
\centering
\includegraphics[width=6.5cm,angle=90]{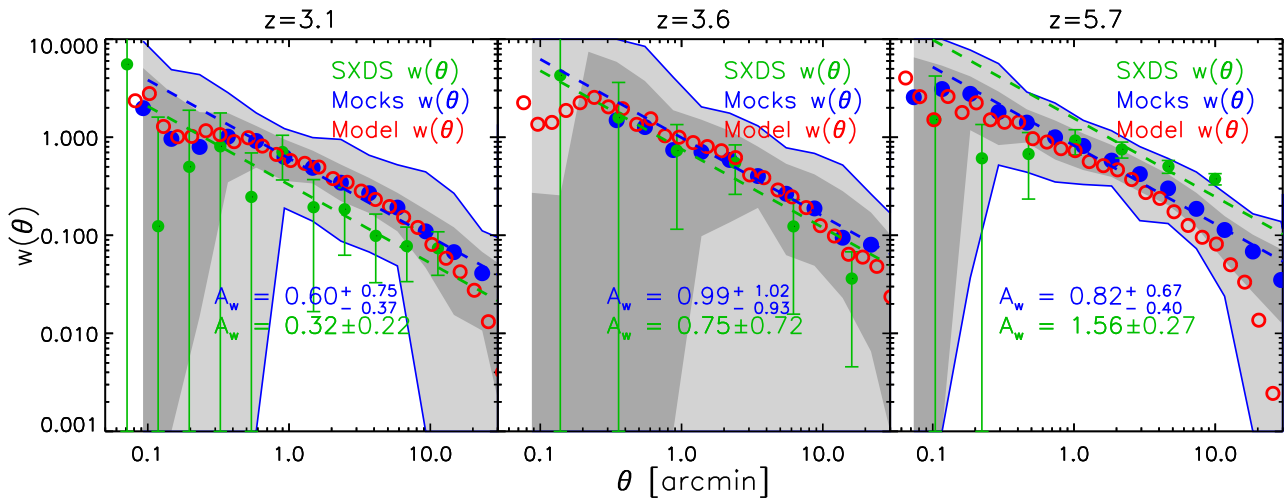}
\caption{Angular correlation functions for the mock SXDS catalogues at
$z=3.1$ \textit{(Left)}, $z=3.6$ \textit{(Center)} and $z=5.7$
\textit{(Right)}.  The blue circles show the median \wtheta\ from the
mock catalogues (after applying the IC correction). The dark and light
grey shaded regions respectively show the 68\% and 95\% ranges of the
distribution of \wtheta\ measured in the mock catalogues. The red open
circles are the Model \wtheta\ calculated using the full simulation
width, averaged over many slices. The green circles show the
observational data from Ouchi et al. The dashed lines show the
power-law fit to the observed \wtheta\ (green) and the median fit to
\wtheta\ from the mock catalogues (blue). The amplitudes $A_w$ of
these fits are also given in the figure. }
\label{fig.w_sxds}
\end{figure*}

We build mock SXDS catalogues following the same procedure as outlined
above.  Fig.~\ref{im.sxds} shows examples of our mock catalogues for
each redshift. As in the previous case, we see that \lya\ emitters on
average trace the higher density regions of the dark matter
distribution. The real surveys have a well defined angular
size. However, the area sampled is slightly different at each
redshift. In order to keep the cross-like shape in our mock catalogues
and be consistent with the exact area surveyed, we scaled the
cross-like shape to cover the same angular area as the real survey at
each redshift.

Fig.~\ref{fig.ng_sxds} shows the distribution of the number of
galaxies in the mock catalogues for the three redshifts surveyed. The
median number of galaxies in the mocks at $z=3.1$ is 316, which
is remarkably similar to the observed number, 356. The 10-90
percentile range of the mocks covers 256--379 galaxies. The
coefficient of variation is $C_v = 0.19$, less than half the value
found for the MUSYC mock catalogues, $C_v = 0.41$.  This reduction is
due mainly to the larger area sampled by the SXDS survey. In the
second slice ($z=3.6$), the redshift is only slightly higher than in
the previous case, but the number of galaxies is much lower. Looking
at the top panel of Fig.~\ref{fig.clf} we see that the observed LFs
are basically the same for these two redshifts. The difference between
the two samples is explained mostly by the different \lya\ flux limits
($1.2 \times 10^{-17} \funits$ for $z=3.1$ and $2.6 \times 10^{-17}
\funits$ for $z=3.6$). For the $z=3.6$ mocks, we find a median number
of 80 and 10-90\% range 60--110, in reasonable agreement
with the observed number of galaxies, 101. The fractional variation in
the number of galaxies, quantified by $C_v = 0.31$, is larger than in
the previous case, due to the smaller number of galaxies. The $z=5.7$
case is similar to the lower redshifts. The median number in the mocks
is 329, with 10-90\% range 255--407, again consistent with
the observed number, 401. The coefficient of variation for this survey
is $C_v = 0.23$, so the sampling variance is intermediate between that
for the $z=3.1$ and $z=3.6$ surveys.

The angular correlation functions of the mock catalogues are compared
with the real data in Fig.~\ref{fig.w_sxds}. The observational data
shown are preliminary angular correlation function measurements in the
three SXDS fields, with errorbars based on bootstrap resampling
(M. Ouchi, private communication).  As in our comparison with the
MUSYC survey, we plot the median correlation function measured from
the mocks, after applying the IC correction, as a representative
\wtheta\ . As before, we also perform a $\chi^2$ fit of a power law to
the \wtheta\ measured in each mock, and to the observed values, to
determine the amplitude $A_w$. The fit is performed over the range $1
< \theta < 10 \arcm$ as before.

The left panel of Fig.~\ref{fig.w_sxds} shows the correlation
functions at $z=3.1$. According to both the error bars on the
observational data, and the scatter in \wtheta\ in the mocks (shown by
the grey shading), this survey provides useful constraints on the
clustering for $1 \la \theta \la 10 \arcm$, but not for smaller or
larger angles, where the scatter becomes very large.  The fitted
amplitude $A_w$ for the observed correlation function is $A_w
=(0.32\pm 0.22)$ (95\% confidence, using the error bars reported by
Ouchi et al.), somewhat below the median value found in the mocks, $A_w
= 0.60$, but within the 95\% range for the mocks ($A_w =
0.23-1.35$). Based on the mocks, the model correlation function is
consistent with the SXDS data at this redshift.

Comparing these results with those we found for the MUSYC survey
(which has a very similar redshift and flux limit to SXDS at $z=3.1$),
we see that the results seem very consistent. The MUSYC survey has a
larger sample variance than SXDS, but the measured clustering
amplitude is very similar in the two surveys.

The middle panel of Fig.~\ref{fig.w_sxds} shows the correlation
function for the $z=3.6$ survey. In this case, the error bars on the
observational data and the scatter in the mocks are both larger, due
to the lower surface density of galaxies in this sample. For the
observed correlation amplitude, we obtain $A_w = 0.75 \pm 0.72$, while
for the mocks we find a median $A_w = 0.99$, with 95\% range
0.06--2.01, entirely consistent with the observational data.

The right panel of Fig.~\ref{fig.w_sxds} shows the correlation
function predictions for $z=5.7$. According to the spread of mock
catalogue results, the \wtheta\ measured here is the most accurate of
the three surveys, due to the large number of galaxies. For the mocks,
we find a median correlation amplitude $A_w = 0.82$, with 95\% range
0.42--1.49. For the observations, we find $A_w= 1.56 \pm 0.27$, if we
assume a slope $\delta=0.8$. The average correlation function in the
mocks agrees well with this slope over the range fitted, but the
observational data for \wtheta\ at this redshift prefer a flatter
slope. The model is however still marginally consistent with the
observational data at 95\% confidence. Similarly flat shapes were also
found in some previous surveys \citep{shimazaku04, hayashino04} in the
same field, but at redshifts $3.1$ and $4.9$ respectively. However,
these surveys are much smaller in terms of area surveyed and number of
galaxies (this is particularly so in \citet{shimazaku04}). This
behaviour in \wtheta\ might be produced by the high density regions
associated with protoclusters in the SXDS fields (M. Ouchi, private
communication), but still this behaviour of \wtheta\ must be confirmed
to prove that it is a real feature of the correlation function.

\subsection{ELVIS Survey}

\begin{figure}
\centering
\includegraphics[width=9cm]{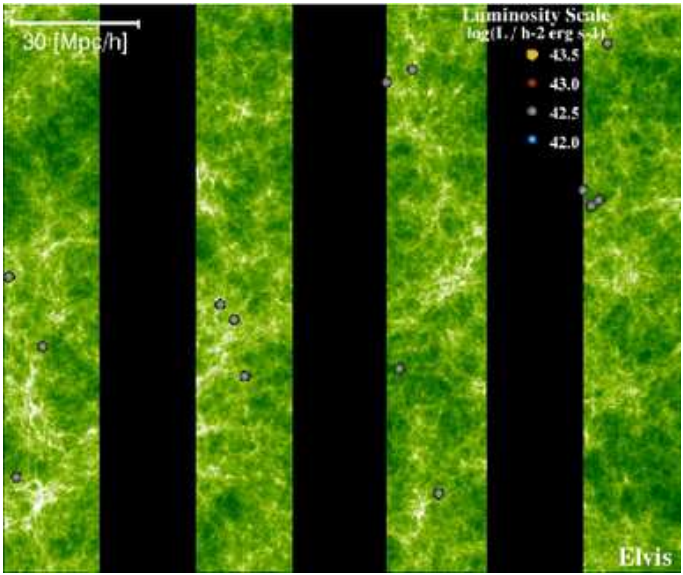}
\caption{An example of a mock catalogue for the ELVIS Survey. The
image shows the observed field of view (four strips). The legend and
colour format are the same as in Figs.\ref{im.musyc} and
\ref{im.sxds}.}
\label{im.elvis}
\end{figure}

One of the main goals of the public surveys planned for the Visible
and Infrared Survey Telescope for Astronomy (VISTA) is to find a
significant sample of very high redshift $(z \sim 8.8)$ \lya\
emitters. This program is called the Emission-Line galaxies with VISTA
survey (ELVIS) \citep{kim07a,kim07b}. The plan for ELVIS is to image
four strips of $11.6 \arcmin \times 68.27 \arcmin$, covering a total area 
of 0.878$\rmn{deg}^2$, as shown in
Fig.~\ref{im.elvis}. This configuration is dictated by the layout of
the VISTA IR camera array. The only current detections of \lya\
emitters at $z>8$ are those of \citet{stark07}, which have not yet
been independently confirmed. \lya\ emitting galaxies at such
redshifts will provide us with valuable insights into the reionization
epoch of the Universe, as well as galaxy formation and evolution.

\begin{figure}
\centering
\includegraphics[width=7cm]{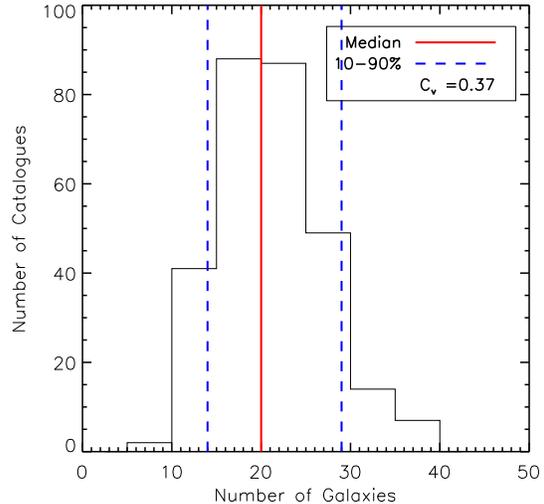}
\caption{Histogram of the number of galaxies in mock catalogues
expected for the ELVIS Survey. The red line shows the median of the
distribution, and the blue dashed lines the 10-90 percentiles of the
distribution.}
\label{fig.ngal_elvis}
\end{figure}

\begin{figure}
\includegraphics[width=8cm]{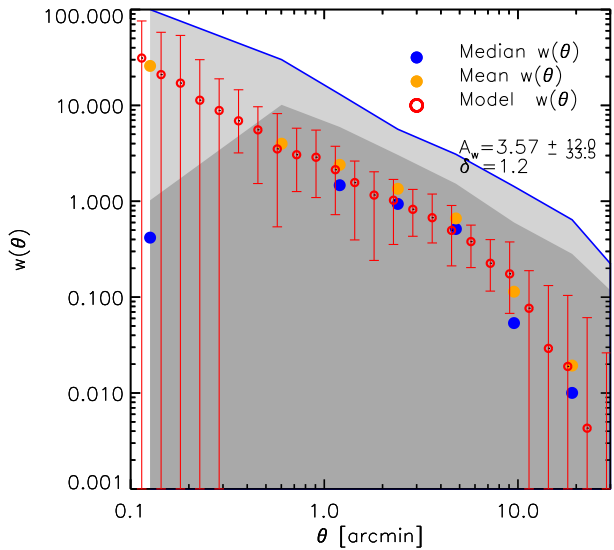}
\caption{Angular Correlation Functions in the mock catalogues of the
ELVIS Survey. The blue circles shows the median \wtheta\ from the mock
catalogues, while the orange circles shows the mean. The dark and
light grey shaded regions respectively show the 68\% and 95\% ranges
of the distribution of \wtheta\ in the mocks.  The red open circles
show the Model \wtheta\, obtained using the full width of the
simulation box. The amplitude and slope of the median power-law fit to
the mocks are also given. }
\label{fig.w_elvis}
\end{figure}

For our mock ELVIS catalogues, we select galaxies with a minimum flux
of $\flya = 3.7 \times 10^{-18} \funits$ and $\ewobs > 100$ \AA{}, as
listed in Table~\ref{table.mock}. (The \ewobs\ limit is just a rough
estimate, although our predictions should not be sensitive to the
exact value chosen.)  Fig.~\ref{im.elvis} shows the expected spatial
distribution of $z=8.5$ galaxies in one of the ELVIS mock
catalogues. Each mock catalogue has four strips, matching the
configuration planned for the real survey. The median number of
galaxies within the mock catalogues is 20, with a 10-90 percentile
spread of 14 to 29 galaxies, as shown in
Fig.~\ref{fig.ngal_elvis}. The fractional variation in number between
different mocks is $C_v=0.37$, which is quite large, but no worse than
for the MUSYC survey at $z=3.1$, even though that survey has 10 times
as many galaxies.

The angular correlation functions of the mock ELVIS catalogues were
calculated in the same way as before (including the integral
constraint correction). Fig.~\ref{fig.w_elvis} shows the median of the
\wtheta\ values measured from each mock catalogue (blue circles), and
also the mean (orange circles). In this case, the distribution of
\wtheta\ values within each angle bin is very skewed, due to the small
number of galaxies in the mock catalogues, and so the mean and median
can differ significantly. The dark and light grey shaded regions show
the ranges containing 68\% and 95\% of the \wtheta\ values from the
mocks, from which it can be seen that the cosmic variance for this
survey is very large. We also show the Model \wtheta\ (red circles),
	which provides our best estimate of the true correlation function
based on the Millennium simulation, and was calculated by averaging
over 10 slices of the simulation, using the full width of the
simulation box. Even here, the error bars on \wtheta\ are quite large,
due to the very low number density of galaxies predicted. We see that
the mean and median \wtheta\ in the mocks lie close to the Model
values for $2\arcm < \theta < 20\arcm$, so in this sense they provide
an unbiased estimate.


The most noticeable feature of Fig.~\ref{fig.w_elvis} is the large
area covered by both the 68\% and 95\% ranges of the distribution of
\wtheta\ in the mocks, which extend down to $w(\theta)=0$. This
indicates that the ELVIS survey will only be able to put a weak upper
limit on the clustering amplitude of $z \sim 8.8$ \lya\ emitters, if
our model is correct. As before, we can quantify this by fitting a
power law to \wtheta\ in our mocks. We notice that the Model \wtheta\
for this sample has a significantly steeper slope, $\delta = 1.2$,
than the canonical value $\delta = 0.8$, and so we do our fits to the
mocks using $\delta = 1.2$. We find a median amplitude in the mocks
$A_w = 3.57^{+12.0}_{-33.5}$, where the error bars give the 95\%
range.

\section{Summary and conclusions}

We have combined a semi-analytical model of galaxy formation with a
high resolution, large volume N-body simulation to make predictions
for the spatial distribution of \lya\ emitters in a $\Lambda$CDM
universe.

Our model for \lya\ emitters is appealingly simple. Using the star
formation history predicted for each galaxy from the semi-analytical
model to compute the production of Lyman continuum photons, we find
that on adopting a constant escape fraction of \lya\ photons the
observed number of \lya\ emitters can be reproduced amazingly well
over a range of redshifts \citep{dell2}. Our modelling of \lya\
emission may appear overly simplistic on first comparison to other
calculations in the literature. For example, \citet{nagamine08}
predicted the clustering of \lya\ emitters in a gas-dynamic
simulation, modelling the \lya\ emission through a \lya\ escape
fraction or a duty cycle scenario.  However, the fraction of active
emitters in the duty cycle scenario needs to be tuned at each
redshift, for which there is no physical justification. Since our
predictions for \lya\ emission are derived from a full model of galaxy
formation, it is straightforward to extract other properties of the
emitters, such as their stellar mass or the mass of their host dark
matter halo (Le Delliou et~al. 2006). In this paper we have extended
this work to include explicit predictions for the spatial and angular
clustering of \lya\ emitters.

We have studied how the clustering strength of \lya\ emitters depends
upon their luminosity as a function of redshift. Generally, we find
that \lya\ emitters show a weak dependence of clustering strength on
luminosity, until the brightest luminosities we consider are reached.
At the present day, \lya\ emitters display weaker clustering than the
dark matter. This changes dramatically at higher redshifts ($z > 3$),
with currently observable \lya\ emitters predicted to be much more
strongly clustered than the dark matter, with the size of the bias
increasing with redshift. We compared the simulation results with
analytical estimates of the bias.  Whilst the analytical results show
the same trends as the simulation results, they do not match well in
detail, and this supports the use of an N-body simulation to study
the clustering.

A key advantage of using semi-analytical modelling is that the
evolution of the galaxy population can be readily traced to the
present day. This gives us some confidence in the star formation
histories predicted by the model. The semi-analytical model passes
tests on the predicted distribution of star formation rates at high
redshift (the number counts and redshifts of galaxies detected by
their sub-millimetre emission and the luminosity function of
Lyman-break galaxies), whilst also giving a reasonable match to the
present day galaxy luminosity function (Baugh et~al. 2005), and also
matching the observed evolution of galaxies in the infrared
\citep{lacey08}. Gas dynamic simulations as a whole struggle to
reproduce the present-day galaxy population, due to a combination of a
limited simulation volume (set by the need to attain a particular mass
resolution) and a tendency to overproduce massive galaxies. The small
box size typically employed in gas dynamic simulations means that
fluctuations on the scale of the box become nonlinear at low
redshifts, and their evolution can no longer be accurately modelled. A
further consequence of the small box size is that predictions for
clustering are limited to small pair separations
(e.g. \citet{nagamine08} use a box of side 33 \mpc, limiting their
clustering predictions to scales $r \la 3 \mpc$). By using a
simulation with a much larger volume than that of any existing \lya\
survey, we can subdivide the simulation box to make many mock
catalogues to assess the impact of sampling fluctuations (including
cosmic variance) on current and future measurements of the clustering
of \lya\ emitters.

We made mock catalogues of \lya\ emitters to compare with the MUSYC
($z=3$) and SXDS ($z=3-6$) surveys, and to make predictions for the
forthcoming ELVIS survey at $z \sim 9$. In the case of MUSYC and SXDS,
we found that the observed number of galaxies lies within the 10-90
percentile interval of the number of \lya\ emitters found in the
mocks. 
We find that high-redshift clustering surveys underestimate their
uncertainties significantly if
they fail to account for cosmic variance in their error budget.
Overall, the measured angular correlation functions are
consistent with the model predictions. The clustering results in our
mock catalogues span a wide range of amplitudes due to the small
volumes sampled by the surveys, which results in a large cosmic
variance. ELVIS will survey \lya\ emitters at very high redshift ($z =
8.8$). Our predictions show that a single pointing will be strongly
affected by sample variance, due to the small volume surveyed and the
strong intrinsic clustering of the \lya\ emitters which will be
detected at this redshift. Many ELVIS pointings will be required to
get a robust clustering measurement.

We have shown that surveys of \lya\ emitters can open up a new window
on the high redshift universe, tracing sites of active star
formation. With increasing redshift, the environments where \lya\
emitters are found in current and planned surveys become increasingly
unusual, sampling the galaxy formation process in regions that are
likely to be proto-clusters and the progenitors of the largest dark
matter structures today. Our calculations show that with such strong
clustering, surveys of \lya\ emitters covering much larger
cosmological volumes are needed in order to minimize cosmic variance
effects.

\section*{Acknowledgements}
AO gratefully acknowledges a STFC Gemini studentship and support of
the European Commission's ALFA-II programme through its funding of the
Latin-American European Network for Astrophysics and Cosmology
(LENAC). CGL acknowledges support from the STFC rolling grant for
extragalactic astronomy and cosmology at Durham.  CMB is supported by
a University Research Fellowship from the Royal Society. { We are very grateful
to Masami Ouchi and collaborators for kindly providing us with their
clustering measurements in advance of publication.} 
We acknowledge a helpful report from the referee.

\section*{ADDENDUM TO PUBLISHED VERSION}
\begin{figure}
\centering
\includegraphics[width=8cm]{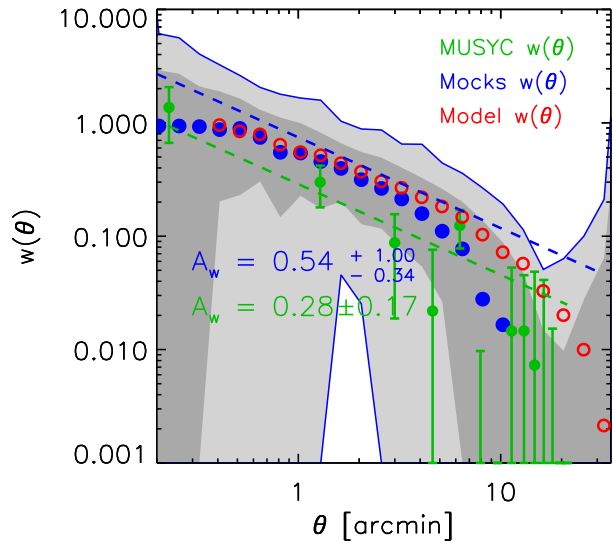}
\caption{ Angular clustering for the MUSYC Survey. Green circles show
\wtheta\ calculated from the observed catalogue \citep{gawiser07}. The
blue circles show the median \wtheta\ from all mock catalogues,
with no correction for the integral constraint effect.  The dark and light grey
shaded regions respectively show the 68\% and 95\% ranges of the
distribution of \wtheta\ measured in the mock catalogues. The red open
circles show the \textit{Model} correlation function, obtained using
the width of the entire simulation box (and the same EW, flux and
redshift limits). The dashed lines show the power-law fit to the
observed \wtheta\ (green) and the median fit to \wtheta\ from the mock
catalogues (blue). The amplitudes $A_w$ of these fits are also given
in the figure.}
\label{fig.wm}
\end{figure}

As explained in the footnote to section 5.1, after the paper was
accepted for publication we learned that the MUSYC datapoints for
\wtheta\ presented in \citet{gawiser07} are in fact not corrected for
the integral constraint (IC) effect.  We therefore show in
Fig.~\ref{fig.wm} the same comparison of angular correlation functions
as was made in Fig. ~\ref{fig.w_musyc}, but now without making the IC
correction in our mock catalgoues.  The effect of not including the IC
correction in the mocks is that there is now much better agreement
between \wtheta\ in the mocks and the MUSYC observational data.

Comparing Fig.\ref{fig.w_musyc} with Fig.\ref{fig.wm} we can also
directly see the effect of the IC correction on \wtheta.  In
Fig.\ref{fig.wm}, the median of the mock catalogue \wtheta\ (blue
circles) and the Model \wtheta (red open circles) agree on small
scales, but differ on scales larger than $2\arcmin$.  Hence, unless
the IC correction is included in the mock catalogues (as is done in
Fig.~\ref{fig.w_musyc}) we get a biased estimate of the {\it true}
clustering. In this case, as shown in Fig.~\ref{fig.w_musyc}, the IC
correction increases the range of agreement with the Model \wtheta\ to
scales up to $\sim 20 \arcmin$.

\end{document}